\documentclass[12pt]{article}
\usepackage{cite,setspace}
\usepackage{amsmath, amsthm, amssymb,upgreek, slashed}

\usepackage{ifpdf}
\ifpdf
  \usepackage[pdftex]{graphicx}
  \usepackage{epstopdf}
\else
  \usepackage[dvips]{graphicx}
\fi
\parskip=\baselineskip
\def\F{{\mathcal F}}
\def\mod{{\mathrm{mod}}}
\def\Jac{{\mathrm{Jac}}}
\def\d{{\mathrm d}}
\def\Z{{\Bbb Z}}
\def\kalpha{\kappa}
\def\kbeta{\lambda}
\def\WCP{{\Bbb{WCP}}}
\def\WW{{\mathfrak B}}
 \def\OO{{\mathrm O}}
\def\RR{{\Bbb R}}
\def\DD{{\mathfrak D}}
\def\Ber{\mathit{Ber}}

\def\SIgma{\Sigma}
\def\C{{\Bbb C}}
\def\D{{\mathcal D}}
\def\R{{\mathcal R}}
\def\veps{\varepsilon}
\def\h{\widehat}
\def\hat{\widehat}
\def\O{{\mathcal O}}
\def\red{{\mathrm{red}}}
\def\OSp{{\mathrm{OSp}}}
\def\MM{{\mathfrak M}}
\def\M{{\mathcal M}}

\def\spin{{\mathrm{spin}}}
\def\Sp{{\mathrm{Sp}}}
\def\SO{{\mathrm{SO}}}
\def\OO{{\mathrm O}}
\def\L{{\mathcal L}}
\def\X{{\mathcal X}}
\def\Y{{\mathcal Y}}
\def\S{{\mathcal S}}
\def\V{{\mathcal V}}
\def\N{{\mathcal N}}
\def\ww{\xi}
\def\J{{\sf J}}
\def\t{\widetilde}
\def\CP{{\Bbb {CP}}}
\def\neg{\negthinspace}
\numberwithin{equation}{section}
\begin{document}
\begin{titlepage}
\begin{center}
{\bf\Large{\hskip.2cm The Super Period Matrix }}
\vskip.15cm
{\bf\Large{With Ramond Punctures}}
\vskip1cm 
{Edward Witten} 
\vskip 1cm
 {\small{\textit{School of Natural Sciences, Institute for Advanced Study,  Princeton NJ USA 08540 \\}}}
\end{center}
\vskip.5cm
\baselineskip 16pt
\begin{abstract}
We generalize the super period matrix of a super Riemann surface to the case that Ramond punctures are present.  
For a super Riemann surface of genus $g$ with $2r$ Ramond punctures, we define, modulo certain choices that generalize those
in the classical theory (and assuming a certain generic condition is satisfied), 
 a $g|r\times g|r$ period  matrix that is
symmetric in the $\Z_2$-graded sense.   As an application, we analyze the genus 2 vacuum amplitude in string theory compactifications to four dimensions
that are supersymmetric at tree level.  We find  an explanation for a result that has been found in orbifold examples in explicit computations by D'Hoker
and Phong: with their integration procedure, the genus 2 vacuum amplitude always vanishes ``pointwise'' after summing over spin structures, and hence is given
entirely by a boundary contribution.  \end{abstract}
\date{November, 2014}
\end{titlepage}
\renewcommand{\baselinestretch}{0.15}\normalsize
\tableofcontents
\renewcommand{\baselinestretch}{1.0}\normalsize 
\section{Introduction}
By analogy with the classical period matrix of an ordinary Riemann surface, one can define\footnote{See \cite{RSV,DPhagain,DPhtwo}
for original references, and section 8 of \cite{Wittensurf} for a review. The last reference also contains a general introduction to super Riemann surfaces.} 
the super period matrix (sometimes just called the period matrix) of a super Riemann
surface.  If $\Sigma$ is a super Riemann surface of genus $g$ with even spin structure, then its super period matrix 
is a $g\times g$ symmetric matrix of positive imaginary part.  Actually, the super period matrix is only
defined for the case that $\Sigma$ has an even spin structure, and even then it is only defined generically:  it can acquire a pole,
with nilpotent residue, when moduli of $\Sigma$ are varied.

The first goal of the present paper is to extend the definition of the super  period matrix to the case of a super Riemann surface
with Ramond punctures.  (A Neveu-Schwarz puncture, which is simply a marked point, does not affect the definition of the super
period matrix.)    It is conceivable that there is more than one reasonable
definition.  The definition we give here
is motivated by an application that we will explain shortly.  In this definition, the super period matrix of a super Riemann surface
$\Sigma$ of genus $g$ with $2r$ Ramond punctures (the number of Ramond punctures is always even)
is a $g|r\times g|r$ matrix, symmetric in the $\Z_2$-graded sense,
whose $g\times g$ bosonic block has positive definite imaginary part.  (The super period matrix is in general not an arbitrary matrix
of this sort, since in general there are Schottky relations.)  Just as in the classical case, the definition of the period matrix depends on 
a choice of $A$-cycles; when (and only when)  Ramond punctures are present, one
has to define fermionic as well as bosonic $A$-cycles.
If one changes the $A$-cycles that are used in defining it, the super period matrix is transformed by an element of
an  integral form of the supergroup $\OSp(2r|2g)$,  generalizing the fact that in the classical case (or for a super Riemann
surface without Ramond punctures), the period matrix is defined up to the action of an integral form of $\Sp(2g)$.   Just as in the absence of Ramond
punctures, the super period matrix is only generically defined, and can acquire singularities as moduli are varied.

The application we have in mind involves superstring perturbation theory in genus 2.
Every $2\times 2$ matrix of positive imaginary part is the period matrix of an ordinary Riemann
surface of genus 2, unique up to isomorphism (Schottky relations only exist in genus $\geq 4$).  Hence, to a super Riemann surface $\Sigma$ of genus 2 with even spin structure, 
we can associate an ordinary Riemann surface $\Sigma_\red$
of the same period matrix.  $\Sigma_\red$ also inherits a spin structure from the spin structure of $\Sigma$, and the association 
$\Sigma\to \Sigma_\red$
gives a natural holomorphic projection $\pi:\MM_{2,+}\to\M_{2,\spin +}$ from the moduli space $\MM_{2,+}$
of super Riemann surfaces $\Sigma$ of genus 2 with even spin structure to its reduced space $\M_{2,\spin+}$ which parametrizes an ordinary Riemann surface
$\SIgma_\red$ with even spin structure.\footnote{The map $\pi$ is everywhere defined (on $\MM_{2,+}$ as opposed to its Deligne-Mumford compactification), in part
because for $g=2$ (unlike $g>2$) the super period matrix has no poles.  Since the odd dimension of $\MM_{2,+}$ is 2,
$\pi$ is actually a splitting of $\MM_{2,+}$.}

 By integrating over the fibers of $\pi$, one can map the two-loop 
vacuum amplitude of superstring theory, which is naturally a measure $\Upsilon$
on $\MM_{2,+}$, to a measure $\pi_*(\Upsilon)$ on $\M_{2,\spin+}$.  This procedure was the starting 
point in the celebrated analysis of the two-loop vacuum amplitude
by D'Hoker and Phong.  (For a review with further references, see \cite{DPh}.
D'Hoker and Phong also went on to calculate scattering amplitudes in genus 2, 
a much more difficult computation that is beyond the
scope of the present paper.) 

To analyze the integral $\int_{\M_{2,\spin+}}\pi_*(\Upsilon)$, it makes sense to first sum over even 
spin structures before performing any integration. In this way, one projects $\pi_*(\Upsilon)$ from $\M_{2,\spin+}$ to $\M_2$, with the spin structure forgotten.
 In their original work, D'Hoker and Phong analyzed this sum over spin structures for superstring theory on $\RR^{10}$ and for 
certain supersymmetric orbifold compactifications
to six dimensions.  They showed that the sum over spin structures vanishes in those models, analogous to the familiar GSO cancellation in genus 1.

Something new happens in general in the case of a compactification to {\it four} dimensions 
that at tree level has $\N=1$ supersymmetry.  (The most simple examples are provided by compactification of the heterotic string
on a Calabi-Yau three-fold.)  In this case, it is possible for a 1-loop effect to generate a Fayet-Iliopoulos $D$-term, triggering the spontaneous
breaking of supersymmetry \cite{DSW,DIS,ADS}.  When this happens, one expects the genus 2 vacuum amplitude to be non-zero and proportional to
$D^2$.  How does this occur in the context of the D'Hoker-Phong procedure for computing the genus 2 vacuum amplitude?

In general \cite{Wittenmore}, the D'Hoker-Phong procedure must be supplemented with a  boundary correction
(a contribution supported on the divisor at infinity in the compactified moduli space).
The boundary contribution to the genus 2 vacuum amplitude vanishes in supersymmetric compactifications above four dimensions,
but in a four-dimensional model with $\N=1$ supersymmetry, it is proportional to $D^2$.

This raises the possibility
 that in such models, the full answer comes from this boundary correction, and that the bulk contribution, computed with
the procedure of D'Hoker and Phong, always vanishes.  Something similar happens in the
same models in one-loop computations of certain supersymmetry-violating mass splittings  \cite{DIS,ADS}.  

In fact, in examples of orbifold compactifications to four dimensions with $\N=1$ supersymmetry
 \cite{DPhlatest}, the same behavior has been found that was found earlier in supersymmetric models above four dimensions:
 the bulk contribution $\pi_*(\Upsilon)$ to the genus 2 vacuum amplitude vanishes after summing over spin structures,
 without any integration over bosonic moduli.  In the present paper, we will use the theory of the
 super period matrix with Ramond punctures to demonstrate that this very striking behavior will occur in all supersymmetric compactifications
 to four or more dimensions.

Perhaps we should remark that general arguments based on
supersymmetric Ward identities (see for example section 4 of \cite{Wittenmore}) can be used to determine the {\it integrated} genus 2 vacuum amplitude, 
but do not explain the  ``{\it pointwise}'' vanishing that  occurs in the  D'Hoker-Phong procedure.  Our goal here is to explain this more detailed phenomenon.
The arguments governing the integrated behavior are completely general and apply for all values of the genus.  The procedure that leads to pointwise
vanishing is defined only for genus $\leq 2$ or at most (as we discuss in section   \ref{twop}) $g\leq 3$.

We define the super period matrix with Ramond punctures and explain some of its simplest properties in sections \ref{odd}-\ref{dimform} of
this paper.  
(A parallel treatment of some of these issues from the point of view of supergravity will appear elsewhere \cite{DPHcurrent}.)
The application to the two-loop vacuum amplitude is in sections \ref{twop}-\ref{lowgenus}.  The general strategy to constrain the vacuum
amplitude via supersymmetry is familiar \cite{Martinec}, and involves comparing the genus $g$ vacuum amplitude to an amplitude computed
 on\footnote{In general, $\MM_{g,n,2r}$ will denote the moduli space of super Riemann surfaces of genus $g$ with $n$ Neveu-Schwarz punctures and $2r$
 Ramond punctures.  We write $\MM_{g,n,2r,\pm}$ if we wish to indicate the type of spin structure.  Similarly, $\M_{g,n}$ is the moduli space of ordinary Riemann
 surfaces of genus $g$ with $n$ punctures, while $\M_{g,n,\spin\pm}$ is the corresponding moduli space with a choice of even or odd spin structure.} $\MM_{g,0,2}$.
  Since we specifically want to constrain the genus 2 vacuum amplitude computed with a procedure that uses the super period matrix, 
 we have to begin with an understanding of the super period matrix of a genus 2
super Riemann surface with 2 Ramond punctures.
The super period matrix in this situation is only generically defined, with singularities on a certain locus in the moduli space.
The trickiest part of our analysis is to show that these singularities do not ruin the argument; see section \ref{lowgenus}.

Some technical issues are treated in appendices.  In Appendix A, we explain in detail why the super period matrix has a pole; in Appendix B, we give an alternative explanation of the fact that the D'Hoker-Phong
procedure for integration over genus 2 supermoduli space requires a correction at infinity; and in Appendix C, we describe some properties of the
 genus 3 analog of the D'Hoker-Phong procedure.

\section{Odd Periods}\label{odd}

We will begin by recalling the definition of a super Riemann surface with or without Ramond punctures. (The reader may want to consult a more detailed reference
such as \cite{RSV} or \cite{Wittensurf}.)  Then we go on to discuss periods.

A super Riemann surface $\SIgma$ is a complex supermanifold of dimension $1|1$ whose tangent bundle $T\SIgma$ is endowed
with a subbundle $\D$ of rank $0|1$ that is completely unintegrable.  Complete unintegrability means that if $s$ is a nonzero section of $\D$,
then $s$ and $\{s,s\}$ are everywhere linearly independent, so that $\{s,s\}$ generates $T\Sigma/\D$.  Thus
$T\Sigma$ fits in an
exact sequence
\begin{equation}\label{morz}0\to \D\to T\Sigma\to \D^2\to 0. \end{equation}
Dually, the cotangent bundle of $\Sigma$ fits in an exact sequence
\begin{equation}\label{porz}0\to \D^{-2}\to T^*\SIgma\to \D^{-1}\to 0.\end{equation}
One can show that locally, one can pick coordinates $z|\theta$ on $\Sigma$ such that $\D$ is generated by
\begin{equation}\label{worz}D_\theta=\frac{\partial}{\partial\theta}+\theta\frac{\partial}{\partial z}.\end{equation}
Such coordinates are called local superconformal coordinates.
Dually the subbundle $\D^{-2}$ of $T^*\SIgma$ is generated by
\begin{equation}\label{norz}\varpi=\d z-\theta\d\theta.\end{equation}

A super Riemann surface $\Sigma$ with Ramond punctures is a complex supermanifold of 
dimension $1|1$ whose tangent bundle
is still endowed with a subbundle $\D$ of rank $0|1$, but now the condition of complete 
unintegrability fails along a certain divisor.
The local behavior is that $\D$ is generated, in some coordinates $z|\theta$, by
\begin{equation}\label{qorz}D^*_\theta=\frac{\partial}{\partial\theta}+\theta z\frac{\partial}{\partial z}. \end{equation}
Since $(D_\theta^*)^2=z\partial_z$, we see that $D_\theta^*$ and $(D_\theta^*)^2$ 
fail to be linearly independent precisely along
the divisor $\F$ defined by $z=0$.  Thus, the exact sequence (\ref{morz}) is replaced by
\begin{equation}\label{oppo}0\to \D\to T\SIgma\to \D^2(\F)\to 0. \end{equation}
Dually, one has
\begin{equation}\label{noppo}0\to \D^{-2}(-\F)\to T^*\Sigma\to \D^{-1}\to 0, \end{equation}
with $\D^{-2}(-\F)$ generated by
\begin{equation}\label{poppo}\varpi^*=\d z-z\theta\d\theta. \end{equation}
More globally, $\Sigma$ may have many divisors $\F_i$ along which the local behavior of the subbundle $\D\subset
T\Sigma$ is as just described.   We call the $\F_i$ Ramond divisors.  If $\Sigma$ is compact,
the number of Ramond divisors is always even.  
What appears in the exact sequences (\ref{oppo}) and (\ref{noppo}) is $\F=\sum_i\F_i$, which we might call the total
Ramond divisor.    Away from $\F$, $\Sigma$ is an ordinary super Riemann surface.  Along $\F$, $\Sigma$ remains
smooth, but there is a singularity in its superconformal structure.

A general holomorphic 1-form on $\Sigma$ can be written locally as $f(z|\theta)\d z+g(z|\theta)\d\theta$.
In contrast to the case of an ordinary Riemann surface, a holomorphic 1-form is not necessarily closed; if
$\mu$ is a holomorphic 1-form, then $\d\mu$ is a holomorphic 2-form, with an expansion
$\d\mu=a(z|\theta)\d z\d\theta+b(z|\theta)(\d\theta)^2$.    In defining differential forms on a supermanifold,
we define the exterior derivative to be odd, so $\d z$ is odd and anticommutes with $\theta$, while $\d\theta$ is
even and commutes with $\theta$ and $\d z$.    

We want to define periods of holomorphic 1-forms. 
Just as in ordinary geometry, periods, as topological invariants, are only defined for  1-forms that are closed.

The most obvious periods are the analogs of classical periods. Let $\mu$ be a  1-form on $\SIgma$.
  If $S$ is an oriented circle and $\alpha:S\to \Sigma$ is any continuous map, then $\mu$ pulls back to an ordinary
  1-form $\alpha^*(\mu)$ on $S$ so we define the integral $\oint_S\alpha^*(\mu)$.  Just as in the classical case,
  if $\mu$ is closed, then $\oint_S\alpha^*(\mu)$ only depends on the homology class determined by the map $\alpha$.
  For our purposes, the case that $\alpha$ is an embedding is sufficient, so we can just think of $S$ as a smooth 
  submanifold\footnote{Though there is apparently not a useful notion of a smooth function on a complex supermanifold, there is a useful notion of a smooth submanifold of a complex supermanifold, or of a continuous map of a smooth submanifold
  to $\Sigma$.  See section 5 of \cite{Wittennotes}.}
  of $\Sigma$, of real dimension 1 (or $1|0$).   In this case, we write just $\oint_S\mu$ rather than $\oint_S\alpha^*(\mu)$.  An embedded circle in the reduced space $\Sigma_\red$
  of $\Sigma$ can be lifted (not canonically, but in a way that is unique up to homology) to an embedded circle in $\Sigma$.  So, as in \cite{RSV}, one can define
  $A$-periods and $B$-periods for  a closed holomorphic 1-form on $\Sigma$ that correspond precisely to the familiar $A$-periods and $B$-periods 
  of a holomorphic differential on $\Sigma_\red$.

Thus if  $\mu$ is  a closed holomorphic 1-form on $\Sigma$,  it has the usual $g$ $A$-periods and $g$ $B$-periods.  However, in the presence
of Ramond punctures,
such a $\mu$ also has what we might call odd periods. As explained above, near any Ramond divisor $\F_0$, we can pick local coordinates $z|\theta$ in which $\F_0$ is defined by $z=0$
and in which  the superconformal structure is defined by
the distribution generated by
\begin{equation}\label{heffo}D_{\theta}^* =\frac{\partial }{\partial\theta} +\theta z \frac{\partial}{\partial z}
\end{equation}
or dually by the subbundle of the contangent bundle generated by
\begin{equation}\label{effo}\varpi^*=\d z-z\theta \d \theta. \end{equation}
In what follows, it is important that $\theta$ is uniquely defined up to
\begin{equation}\label{ffoo}\theta\to \pm (\theta+c)~~\mod~z,\end{equation}
where $c$ is an odd constant.\footnote{Since $z|\theta\to z| -\theta$ preserves $\varpi^*$, it is obvious that the superconformal
structure determines $\theta$ only 
up to sign.  To see that $\theta$ is also only defined up to a shift $\theta\to\theta+c$ with $c$ an odd constant, one observes
that the supergroup of dimension $0|1$ that acts by $z|\theta\to z(1-c\theta)|\theta+c$ preserves the superconformal structure of $\Sigma$
and acts transitively on $\F_0$.  One can verify that along $\F_0$, any superconformal transformation is equivalent to 
$z|\theta\to z(1-c\theta)|\pm(\theta+c)$.}  In particular, at $z=0$, we are not free to rescale $\theta$ except by $\pm 1$, since this would
disturb the relation between the two terms in $\varpi^*$ or in $D_\theta^*$.  Accordingly, the 1-form $\d\theta$ on $\F_0$ is well-defined
up to sign.  We consider a choice of what we mean by $\d\theta$ as opposed to $-\d\theta$ to represent an ``orientation'' of $\F_0$.

The odd periods of a closed holomorphic 1-form $\mu$ are now defined as follows. Since $\F_0$ is defined by $z=0$, when restricted
to $\F_0$, we have
\begin{equation}\label{doddo}\mu=\frac{w}{\sqrt{2\pi\sqrt{-1}} }\,\d \theta~\mod~z.\end{equation}
  Here $w$ is a constant -- independent of $\theta$ -- since
$\d\mu=0$. 
(The odd-looking factor $\sqrt{2\pi\sqrt{-1}}$ in the denominator will be convenient later.)
 We simply define $w$ to be the odd period of $\mu$ associated to the Ramond divisor $\F_0$. Thus, an odd ``period''
is not defined by an integral but by ``evaluation'' of a 1-form along a Ramond divisor.  The sign of the odd period depends on the orientation
of the Ramond divisor, somewhat analogously to the fact that a 1-manifold $\gamma$ in an ordinary Riemann surface $\Sigma_0$ must
be oriented if one wishes to define the sign of the period $\oint_\gamma\mu$ of a closed 1-form $\mu$.
With $2r$ Ramond
divisors $\F_1,\dots,\F_{2r}$, we define in this way $2r$ odd periods $w_1,\dots, w_{2r}$. 

Thus in all, on a super Riemann surface $\Sigma$ of genus $g$ with $2r$ Ramond punctures, we define $2g$ even periods, just as in the
classical theory, and $2r$ odd periods.    What is the natural symmetry group acting on these periods?  On the bosonic periods,
we have the usual group $\Sp(2g;\Z)$ of automorphisms of $H^1(\SIgma;\Z)$ (which coincides with $H^1(\SIgma_\red;\Z)$).  We can think of $\Sp(2g;\Z)$ as the automorphism group
of the skew form $\sum_{i=1}^g \d a^i\wedge \d b_i$ defined over $\Z$, where $a^i,b_j$, $i,j=1,\dots,g$ are even variables so $\d a^i,$ $\d b_j$ are odd. The natural symmetries
 acting on the  fermionic periods are as follows:  we could permute the Ramond divisors, thus permuting the
 fermionic periods, or we could reverse the orientations of the Ramond divisors and thereby reverse the signs
 of the fermionic periods.  The permutations and sign changes
make up a finite group with $2^{2r}(2r)!$ elements.  We can think of this group as a form of $\OO(2r;\Z)$, since it is the automorphism group of the quadratic form $\sum_{k=1}^{2r}\d x_k^2$, defined
over $\Z$, where the $\d x_k$ are the even differentials of odd variables $x_k$, $k=1,\dots,2r$.  There are no obvious symmetries between even and odd periods. If we combine the two constructions, the full group $\Sp(2g;\Z)\times \OO(2r;\Z)$
that acts on the even and odd periods is the automorphism group of a form
\begin{equation}\label{tomega}\Theta=\sum_i \d a^i\wedge \d b_i - \sum_k \d x_k^2  \end{equation}
that is symmetric in the $\Z_2$ graded sense in $2g$ odd variables $\d a^i$, $ \d b_j$, and $2r$ even variables $\d x_k$.  (For now
the sign of the $\sum_k \d x_k^2$ term is an arbitrary choice.) We think of
this as a sort of superanalog of the intersection form of an ordinary Riemann surface.  Over $\Z$,
there are no symmetries that exchange even and odd variables, so $\Sp(2g;\Z)\times \OO(2r;\Z)$ can be interpreted as
 $\OSp(2r|2g;\Z)$, that is, as a form over $\Z$ of the orthosymplectic group.  
 If we work over $\RR$, of course the supergroup $\OSp(2r|2g;\RR)$ of symmetries of the form $\Theta$
does  mix even and odd variables.  

Typically, not all of $\OSp(2r|2g;\Z)$ is realized as symmetries
 in string theory.  A super Riemann surface has a spin structure, so usually one has 
to consider only the subgroup
of $\Sp(2g ;\Z)$ that preserves a spin structure.  Also it usually is more useful to consider the Ramond divisors to be labeled (or distinguishable), in which
case one considers only the sign changes rather than permutations in $\OO(2r;\Z)$.  Finally, in a certain sense, the space spanned
by $x_1,\dots,x_{2r}$ has a natural orientation, as we explain in section \ref{anex}, so one can replace $\OO(2r;\Z)$ by $\SO(2r;\Z)$.

\section{Closed Holomorphic Two-Forms}\label{closed}

Let us recall how one proves the symmetry of the period matrix of an ordinary Riemann surface $\Sigma_0$.  If $\mu$ and $\mu'$ are
closed 1-forms on $\Sigma_0$, one has the topological fact
\begin{equation}\label{zelmo}\int_{\Sigma_0}\mu\wedge \mu'=\sum_{i=1}^g\left(\oint_{A^i}\mu\oint_{B_i}\mu'-\oint_{A^i}\mu'\oint_{B_i}\mu \right).
 \end{equation}   
If $\mu$ and $\mu'$ are holomorphic 1-forms, then $\mu\wedge\mu'$, which would be a holomorphic 2-form, vanishes identically.
So the left hand side of (\ref{zelmo}) vanishes, and this leads to the symmetry of the period matrix.  

To imitate this argument on a super Riemann surface $\Sigma$, we face two difficulties:  {\it (i)} if we  view $\Sigma$ as a smooth
supermanifold (this can be done, though not quite in a canonical way), then it has dimension $2|1$, but a two-form $\mu\wedge\mu'$
can only be integrated on a manifold of dimension $2|0$; {\it (ii)} on a super Riemann surface, a holomorphic 2-form is not necessarily 0.

The resolution of the first point is simply that, as explained in \cite{RSV}, if $\Sigma$ is a super Riemann surface, with reduced space
$\SIgma_\red$, then $\SIgma_\red$ can be embedded in $\Sigma$ in a way that is not canonical (or holomorphic)
but is unique up to homology.  The image of the embedding is a smooth submanifold $\Sigma^*\subset \Sigma$ of dimension $2|0$, and the
proof of symmetry of the period matrix is made using integrals over $\Sigma^*$ rather than $\Sigma$. 

The resolution of the second point was also explained in \cite{RSV} (see also section 8.2 of \cite{Wittensurf}).  On a super Riemann surface (without Ramond punctures), a closed
holomorphic 2-form is exact, so if $\mu$ and $\mu'$ are holomorphic 1-forms, then $\mu\wedge\mu'=\d\lambda$ for some $\lambda$.
This implies vanishing of $\int_{\Sigma^*}\mu\wedge\mu'$, just as if $\mu\wedge\mu'$ were 0, and leads to the proof of symmetry of the
super period matrix.

Before describing these arguments, and explaining how they must be modified in the presence of Ramond punctures, we give
a more elementary example.  Let $W$ be the supermanifold of dimension $1|1$ defined as the quotient
of $\C^{1|1}$ by
\begin{align}\label{torg}z|\theta& \to z+1|\theta+\alpha\cr
                                     z|\theta& \to z+\sqrt{-1}|\theta. \end{align}
 Here $\alpha$ is an odd constant.  $W$ is a complex supermanifold of dimension $1|1$, but not a super Riemann surface since the identifications in eqn. (\ref{torg})
 do not preserve a superconformal structure.  The expression $\d z\,\d\theta$ defines a closed holomorphic 2-form on $W$ that is 
 globally-defined and is not
 exact (we can write it locally as $\d(\theta\,\d z)$, but $\theta\,\d z$ is not globally-defined, since it is not invariant under $z|\theta\to z+1|\theta+\alpha$).  The reduced space $W_\red$ of $W$ is
 the ordinary Riemann surface of genus 1 defined by $z\cong z+1\cong z+\sqrt{-1}$.  It  
 can be embedded in $W$ as the submanifold $W^*$ defined by the equation $\theta=\alpha\,\mathrm{Re} \,z$.    Then one finds
 that $\int_{W^*}\d z\,\d\theta = -\alpha\sqrt{-1}\not=0. $   The nonzero value of the integral is one way to prove that $\d z\,\d\theta$ is not exact.
 (Since $\d z\,\d\theta$ is closed, the value of this integral does not depend on the
 precise choice of $W^*$.)

On a super Riemann surface $\SIgma$ without Ramond punctures, the proof that a closed holomorphic 2-form $\Psi$
is exact proceeds as follows.
In local superconformal coordinates $z|\theta$,  $\Psi$ can be expanded
\begin{equation}\label{zug} \Psi=(\d\theta)^2 \,p(z|\theta)+\d\theta\, \varpi \,\rho(z|\theta), \end{equation}
where as usual $\varpi=\d z-\theta\d\theta$.  The condition $\d\Psi=0$ gives
\begin{equation}\label{wug} \rho=D_\theta p,\end{equation}
and then one finds that 
\begin{equation}\label{lug} \Psi=\d f,~~~ f=-\varpi \,p(z|\theta). \end{equation}
Even though we have computed in local superconformal coordinates, the object $f$ that we have defined
does not depend on this choice.  This statement can be explained in the following way (as in footnote 34 of \cite{Wittensurf}), using
the exact sequence $0\to \D^{-2}\to T^*\SIgma\to\D^{-1}$.  The projection $T^*\SIgma\to\D^{-1}$, tensored with itself,
gives a holomorphic map $\wedge^2T^*\SIgma\to \D^{-2}$.  When this is composed with the inclusion $\D^{-2}\to T^*\Sigma$,
we get a natural map $\wedge^2 T^*\SIgma\to T^*\SIgma$ which in local superconformal coordinates is the map $\Psi\to \varpi p=-f$.

Let us now see how these considerations are modified in the presence of a Ramond divisor.  As usual, we consider a local
model with coordinates $z|\theta$ and a superconformal structure defined by $\varpi^*=\d z -z \theta\d\theta$.  The Ramond
divisor $\F$ is defined by $z=0$.   A holomorphic 2-form $\Psi$ can be expanded 
\begin{equation}\label{dolf}\Psi=(\d\theta)^2p(z|\theta)+\d\theta\varpi^* \rho(z|\theta). \end{equation}
The condition $\d\Psi=0$ implies that $p(z|\theta)$ is independent of $\theta$ at $z=0$, so
we define the constant
\begin{equation}\label{ug}p =p(0|0). \end{equation}
Note that $p$ is  completely well-defined, as it is not affected by the indeterminacy (\ref{ffoo}) of $\theta$.  

Away from $z=0$, we can replace $z|\theta$ by local superconformal coordinates $z|\h\theta=z|\theta
z^{1/2}$ (these are superconformal coordinates, because in these coordinates the usual expression $\varpi^*=\d z-z\theta\d\theta$ near a Ramond puncture
 takes the standard superconformal form $\d z-\h\theta\d\h\theta$). In these coordinates, $\Psi$ has
 a pole at $z=0$, with
  $\Psi\sim(\d\h\theta)^2 p/z$, and hence $f$ as determined in eqn. (\ref{lug})
behaves as
\begin{equation}\label{grug} f\sim -\frac{p\varpi^*}{z}+\O(1), ~~z\to 0. \end{equation}
Going back to  the coordinates $z,\theta$ that behave well near the Ramond divisor, since $\varpi^*=\d z-z\theta\d\theta$,
we have simply
\begin{equation}\label{burg} f\sim -p\frac{\d z}{z}+\O(1),~~z\to 0. \end{equation}
Away from $z=0$, it is still true that $\Psi=\d f$.  But because of the pole of $f$ at $z=0$, there is actually a delta-function
contribution to $\d f$ at $z=0$: $\d f=\Psi-2\pi\sqrt{-1}p\delta$, where $\delta$ is a two-form delta function that is Poincar\'e
dual to the Ramond divisor $\F$  at $z=0$.    Including many Ramond divisors $\F_\alpha$, we get 
\begin{equation}\label{morog} \d f=\Psi-2\pi \sqrt{-1}\sum_\alpha p_\alpha \delta_\alpha, \end{equation}
where $\delta_\alpha$ is dual to $\F_\alpha$ and $-p_\alpha$ is the residue of the corresponding pole in $f$.

Integrating this formula over a submanifold $\Sigma^*\subset\SIgma$ that is isomorphic to $\Sigma_\red$, we get
\begin{equation}\label{norog}\int_{\Sigma^*}\Psi =2\pi\sqrt{-1}\sum_\alpha p_\alpha. \end{equation}

\section{The Super Period Matrix}\label{superp}

If  $\Psi=\mu\wedge\mu'$, where $\mu$ and $\mu'$ are two closed holomorphic 1-forms on a super Riemann surface $\SIgma$
with Ramond punctures, then we can combine (\ref{norog})  with the topological formula
(\ref{zelmo}), with the result that
\begin{equation}\label{zelmof}\sum_{i=1}^g\left(\oint_{A^i}\mu\oint_{B_i}\mu'-\oint_{A^i}\mu'\oint_{B_i}\mu \right)=2\pi \sqrt{-1}\sum_\alpha
p_\alpha.
 \end{equation}   
 
 To make use of this, we must express the constants $p_\alpha$ in terms of the odd periods $w_\alpha$ and $w'_\alpha$ of
 $\mu$ and $\mu'$.   Recalling that the odd period is simply the constant $w$ in eqn. (\ref{doddo}), we see that
 near the Ramond divisor $\F_\alpha$, we have $\mu\sim w_\alpha \d\theta/\sqrt{2\pi\sqrt{-1}}$, $\mu'\sim w'_\alpha\d\theta/\sqrt{2\pi\sqrt{-1}}$,
 and hence $\mu\wedge\mu'\sim (\d\theta)^2w_\alpha w_\alpha'/2\pi\sqrt{-1}$.  
 Thus $2\pi\sqrt{-1}p_\alpha=w_\alpha w'_\alpha$.

If then we denote the $A$- and $B$-periods as $a_i=\oint_{A^i}\mu$, $b^i=\oint_{B_i}\mu$ and similarly $a'_i=\oint_{A^i}\mu'$,
$b'^i=\oint_{B_i}\mu'$, then we arrive at the analog of the Riemann bilinear relations for a super Riemann surface with Ramond punctures:
\begin{equation}\label{bilrel}\sum_{i=1}^g\left(a_i b'^j-a'_ib^j\right)-\sum_{\alpha=1}^{2r} w_\alpha w'_\alpha = 0. \end{equation}

On the right hand side of eqn. (\ref{bilrel}), we see the ``intersection form'' $\Theta$ that was introduced in eqn. (\ref{tomega}).  Thus,
it is natural to introduce a space $\Lambda\cong \C^{2g|2r}$ that is endowed with this quadratic form (tensored with $\C$).  We denote
the intersection form  on $\Lambda$ as $\langle~,~\rangle$.
We combine the whole collection of even and odd periods of $\mu$ to a vector $\upmu\in\Lambda$:
\begin{equation}\label{dormo}\upmu=\{a_i,b^j|w_\alpha\},~~i,j=1,\dots g, ~\alpha=1,\dots, 2r.\end{equation}
Similarly, the periods of $\mu'$ combine to $\upmu'=\{a'_i,b^{'j}|w'_\alpha\}\in \Lambda$.
Eqn. (\ref{bilrel}) is equivalent to  $\langle\upmu,\upmu'\rangle=0$. 

 In other words, by analogy with the classical case, the bilinear
relations assert that the subspace $\Lambda_0\subset \Lambda$ that is spanned by the periods of holomorphic 1-forms is
an isotropic subspace: the bilinear form $\langle~,~\rangle$ vanishes when restricted to $\Lambda_0$.
In section \ref{dimform}, we will show that generically, for $r>0$ (or for $r=0$ with even spin structure) the space of closed holomorphic 1-forms on $\Sigma$ has dimension
$g|r$. In this case, $\Lambda_0$ is middle-dimensional in $\Lambda$, and thus it is
a maximal isotropic subspace of $\Lambda$, again by analogy with the classical case.

Just as in the more familiar case $r=0$, the information about a maximal isotropic subspace of $\Lambda$ can generically
be encoded by a super period matrix.  The super period matrix
will now be a $g|r\times g|r$ matrix that will be symmetric in the $\Z_2$-graded sense, as described more concretely below.
To define the classical period matrix of an ordinary Riemann surface $\Sigma_0$, one starts by picking a set of $A$-periods.
This amounts to picking a maximal isotropic subspace -- of a particularly simple and convenient sort -- 
for the intersection form on $H^1(\Sigma_0,\Z)$.   To generalize this for a super Riemann surface with Ramond punctures,
we similarly must first pick a simple maximal isotropic subspace for the  form $\Theta$, which we regard as the superanalog of the
classical intersection form.   We again use a set of $A$-periods as a maximal set of even null vectors, but what is a natural set
of odd null vectors?  
The simplest choice seems to be to  order the fermionic periods as $w_1,w_2,\dots,w_{2r}$
and then form the complex linear combinations \begin{equation}\label{melz}w^{\zeta}=\frac{1}{\sqrt 2}(w_{2\zeta-1}+\sqrt{-1}w_{2\zeta}),~~
\zeta=1,\dots ,r.\end{equation}  The complementary
fermionic periods are \begin{equation}\label{elz}\t w_\zeta=\frac{1}{\sqrt 2}(w_{2\zeta-1}-\sqrt{-1}w_{2\zeta)},~~\zeta=1,\dots,r.\end{equation}  
The $w^{\zeta}$ and $\t w_\zeta$ will be the
fermionic analogs of $A$-periods and $B$-periods.  

Now we define a basis of closed holomorphic 1-forms $\sigma_1,\dots,\sigma_g|\nu_1,\dots,\nu_r$ by requiring
\begin{equation}\label{zorb}a^i(\sigma_j)=\delta^i_j,~~ w^\eta(\sigma_j)=0 \end{equation}
and
\begin{equation}\label{omorb} a^i(\nu_\zeta)=0,~~w^{\eta}(\nu_\zeta)=\delta^\eta_\zeta. \end{equation}
 For this definition to make sense, the choice of fermionic $A$-periods must be generic enough
so that closed holomorphic 1-forms obeying the conditions (\ref{zorb}) and (\ref{omorb}) exist and are unique.   
The condition for this is that if $\Lambda'$ is the subspace of $\Lambda$ defined by $w^\zeta=0=a^i$,
then we must have $\Lambda'\cap \Lambda_0=0$.  
We return to this condition in sections \ref{nonperiods} and  \ref{anex}, and for now just remark that it places a non-trivial constraint on
the choices of ordering and signs in the definition of the $w^\zeta$.

Finally, we define the super period matrix $\h \Omega$ by specifying its matrix elements
\begin{align}\label{motto}\h\Omega_{ij}&=\oint_{B^j}\sigma_i=b_j(\sigma_i) \cr
\h\Omega_{i\eta}&=\t w_{\eta}(\sigma_i)\cr \h\Omega_{\eta j}&=\oint_{B^j}\nu_\eta=b_j(\nu_\eta)\cr
 \h\Omega_{\eta\zeta}& = \t w_{\zeta}(\nu_\eta)                  .\cr \end{align} 
 As in the classical theory, the bilinear relations (\ref{bilrel})
  imply that $\h\Omega$ is symmetric in the $\Z_2$-graded sense: $\h\Omega_{ij}=\h\Omega_{ji}$,
 $\h\Omega_{i\eta}=\h\Omega_{\eta i}$, $\h\Omega_{\eta\zeta}=-\h\Omega_{\zeta\eta}$.
If we  write the super period matrix in blocks
 \begin{equation}\label{myr}\begin{pmatrix} g\times g & g\times r\cr r\times g & r \times r\end{pmatrix}\end{equation} 
 then the upper left $g\times g$ block, which we will call $\h\Omega_{g\times g}$, corresponds most closely to the ordinary period matrix in the classical theory of Riemann
 surfaces.  We will call this the pseudoclassical block.  
 If reduced modulo odd variables, it coincides with the ordinary period matrix of the reduced space $\SIgma_\red$.
 This will be clear in section \ref{anex}.  So in particular $\h\Omega_{g\times g}$ has positive-definite imaginary part. 
   Note that the pseudoclassical block
depends on the choice of fermionic $A$-periods (though this dependence disappears if we reduce 
modulo the odd variables), since the definition of the $\sigma_i$ depends on that choice.

\section{The Space Of Closed Holomorphic One-Forms}\label{dimform}

\subsection{Counting Closed Holomorphic One-Forms}\label{counting}

To count the closed holomorphic 1-forms on a genus $g$ super Riemann surface
$\Sigma$, first assume that $\SIgma$ is split, with a reduced space $\Sigma_\red$
whose canonical bundle and tangent bundle we denote as $K$ and $T$.  To begin with, assume there are no Ramond punctures,
Then
$\SIgma$ can be constructed as the total space of an odd line bundle $\varPi T^{1/2}\to \Sigma_\red$.  
Here $T^{1/2}$ is a square root of $T$ corresponding to a choice of spin structure, the inverse of $T^{1/2}$ will be denoted $K^{1/2}$,
and for a line bundle $\L$, $\varPi \L$ is $\L$ with the fiber understood to be odd.  There is a natural
projection $\pi:\Sigma\to \Sigma_\red$ (and an embedding of $\Sigma_\red$ in $\SIgma$ as the zero-section of $\varPi T^{1/2}$).

There is always a $g$-dimensional space of holomorphic 1-forms $b(z)\d z$ on $\SIgma_\red$.  These can be pulled back via
$\pi$ to closed holomorphic 1-forms on $\SIgma$.     There actually are additional odd holomorphic 1-forms 
$b(z)\theta\d\theta$, but they are not closed.
The situation for even closed holomorphic 1-forms is more interesting. 
 An even  holomorphic 1-form
is in general  $\lambda= a(z)\theta\d z + c(z)\d\theta$, but for $\lambda$ to be closed, this expression must reduce to $\lambda=\d(c(z)\theta)$.
Here in classical geometry $c(z)$ represents a holomorphic section of $K^{1/2}$, that is, an element of $H^0(\SIgma_\red,K^{1/2})$.
Generically, if the spin structure of $\Sigma$ is even, $H^0(\Sigma_\red,K^{1/2})=0$.  In this case the space of closed holomorphic
1-forms has dimension $g|0$ (we will reverse the parity in writing dimension formulas). Their periods are used to define a $g\times g$ super period matrix $\hat\Omega_{ij}$, which is symmetric
and has positive definite imaginary part, just as in the classical case.

Again assuming that $\Sigma$ has even spin structure, in genus $g\geq 3$, there is a divisor  $\DD\subset \M_{g,\spin+}$ along which $H^0(\SIgma_\red,
K^{1/2})\not=0$.  The space of closed holomorphic 1-forms is then of dimension $g|s$, for some (even) $s>0$.  There are more closed holomorphic
1-forms than periods so some closed holomorphic 1-forms must have vanishing periods.  In fact, as explained in the last paragraph, the
even closed holomorphic 1-forms are exact ($\lambda=\d(c(z)\theta)$), so their periods vanish. In defining a super period matrix, one can take
the quotient of the space of closed holomorphic 1-forms by the subspace consisting of those whose periods vanish.  For $\Sigma$ split, the
quotient space always
has dimension $g|0$.  
So as long as $\Sigma$ is split, the condition $H^0(\Sigma_\red,K^{1/2})\not=0$ does not lead to trouble in defining the super period
matrix.
 
 To define the super period matrix without assuming that $\Sigma$ is split, we need to know that (away from $\DD$) the space of closed holomorphic 1-forms is still of dimension 
 $g|0$ when
the odd moduli of $\SIgma$ are introduced.  This is true but not completely trivial.   One elegant proof\footnote{This proof
was given in \cite{RSV}.  See also
\cite{Wittensurf}, section 8 and Appendix D, for a detailed explanation.} 
uses the fact that there is a natural 1-1 correspondence between closed holomorphic
1-forms and holomorphic sections of $\Ber(\SIgma)$, the Berezinian line bundle of $\Sigma$.   The correspondence is given
by an explicit formula; in local superconformal coordinates, a holomorphic section $\phi(z|\theta)[\d z|\d\theta]$ of $\Ber(\Sigma)$
corresponds to the closed holomorphic 1-form  $\mu=\d\theta\phi+\varpi D_\theta\phi$.   To show that the space of holomorphic 1-forms
on $\SIgma$ is still of dimension $g|0$ when $\Sigma$ is not split, one must show the 
analogous statement for $H^0(\Sigma;\Ber(\SIgma))$:
its dimension should not jump when odd moduli are turned on.  On general grounds,
this is true if and only if $H^1(\Sigma,\Ber(\Sigma))$ varies as the fiber of a locally-free sheaf (or vector bundle).  
But $H^1(\SIgma,\Ber(\SIgma))$
is Serre-dual to $H^0(\SIgma,\O)$, where $\O$ is the sheaf of holomorphic sections on $\Sigma$.  Away from the divisor
$\DD$, one has $H^0(\Sigma,\O)\cong\C$, generated by the constant function 1.  In particular, $H^0(\SIgma,\O)$ is locally-free,
and hence
also are $H^1(\SIgma,\Ber(\SIgma))$ and $H^0(\SIgma,\Ber(\SIgma))$.

This reasoning fails along the divisor $\DD\subset \M_{g,\spin+}$, because given $c\in H^0(\SIgma_\red,K^{1/2})$, there is an odd holomorphic
function $c(z)\theta$ on $\Sigma$. Thus for a split super Riemann surface $\Sigma$, vanishing of $H^0(\SIgma_\red,K^{1/2})$ is a necessary
and sufficient condition for $H^0(\Sigma,\O)\cong \C$.
 What actually happens near $\DD$ is that, although the super period
matrix is well-defined and holomorphic as long as the odd moduli vanish, or in other words along the split locus $\M_{g,\spin+}\subset \MM_{g,+}$,
it develops a pole (with nilpotent residue) as soon as one varies away from that locus.  This follows from the formula of D'Hoker and Phong
\cite{DPhagain} for the dependence of the super period matrix on odd moduli.  (See section 8.3 of \cite{Wittensurf}, or Appendix \ref{perturb}
below.)

Now let us consider the case that $\SIgma$ is a super Riemann surface with  Ramond punctures.  
Again we start with the split case.

We pick in the reduced space $\Sigma_\red $ of $\SIgma$
a collection of distinct points
$p_1,\dots, p_{2r}\in\Sigma_\red$ that will represent the Ramond punctures, and a line bundle $\R$ endowed with an
isomorphism\footnote{\label{zep} Such a line bundle defines 
what we call a generalized spin structure.  One is free to tensor $\R$ with a line bundle of order
2, so for any $r$ and any points $p_1,\dots,p_{2r}$, there  are $2^{2g}$ generalized
spin structures.  For $r=0$, the choice of $\R$ is tantamount to an ordinary spin structure on 
$\Sigma$;  the spin structures on $\Sigma$ can be naturally
divided into
odd and even ones.  For $r>0$, the $2^{2g}$ generalized spin structures are permuted transitively under
monodromy of the points $p_i$, so there is no notion of an even or odd generalized spin structure.}
\begin{equation}\label{elmonk}\R^2\cong T\otimes \O(-p_1-\dots -p_{2r}) \end{equation}
or equivalently
\begin{equation}\label{zelmonk} K\otimes \R\cong \R^{-1}(-p_1-\dots -p_{2r}). \end{equation}
The line bundle $\R$ has degree $1-g-r$.
$\Sigma$ is then defined to be the total space of the line bundle $\varPi\R\to\Sigma_\red$. As before, there are
 projections $\pi:\Sigma\to\Sigma_\red$ and an embedding $\Sigma_\red\subset\Sigma$.
 Away from the points $p_1,\dots, p_{2r}$, the line bundle $\R$ is a 
square root of $T$
and $\Sigma$ is an ordinary super Riemann surface, which can be described by local superconformal coordinates $z|\theta$
and superconformal structure generated by $D_\theta=\partial_\theta+\theta\partial_z$.   However, because the isomorphism
$\R^2\cong T$ is only valid away from the points $p_i$, the superconformal structure breaks down 
along the  divisors $\F_\alpha=\pi^{-1}(p_\alpha)\subset \Sigma$.  Those divisors are Ramond divisors,
representing singularities in the superconformal structure of $\Sigma$. 

Closed holomorphic 1-forms can be described as before.  Holomorphic 1-forms on $\Sigma_\red$ can be pulled back to give
a $g|0$-dimensional space of closed holomorphic 1-forms on $\Sigma$.  As before, these are the only odd closed holomorphic
1-forms, and even ones are of the form
$\d(a(z)\theta)$, where now geometrically $a(z)$ is a holomorphic section of $\R^{-1}\to\Sigma_\red$.  This line bundle is of degree $g-1+r$,
so generically $H^0(\SIgma_\red,\R^{-1})$ is of dimension $r$ and $H^1(\SIgma_\red,\R^{-1})=0$. 
This fails on a locus $\WW$ characterized by $H^1(\Sigma_\red,\R^{-1})\not=0$.  This locus, which will be studied in section
\ref{badset},  is a rough analog of the
theta-null divisor $\DD$ for $r=0$.

Overall, away from  $\WW$, the space of closed holomorphic 1-forms on a split super Riemann surface $\Sigma$
has dimension $g|r$, as assumed in section
\ref{superp} in defining the super period matrix.  To show that this remains so if $\SIgma$ is not assumed to be split, one can
adapt the arguments that are used in the absence of Ramond punctures.    In the presence of Ramond punctures, closed holomorphic
1-forms correspond (see Appendix D.1 of \cite{Wittensurf}) not to elements of $H^0(\Sigma,\Ber(\Sigma))$, but to elements of $H^0(\Sigma,
\Ber'(\Sigma))$, where a section of $\Ber'(\SIgma)$ is a 
section of $\Ber(\SIgma)$ that is allowed to have a simple pole, with $\theta$-independent
residue, along a Ramond divisor.  The space of closed holomorphic 1-forms is locally-free if $H^1(\Sigma,\Ber'(\SIgma))$ is locally-free.  By
Serre duality, this is equivalent to $H^0(\Sigma,\O')$ being locally-free, where $\O'$ is the sheaf of holomorphic functions on $\SIgma$ that are constant
when restricted to a Ramond divisor.  Equivalently, $H^0(\SIgma,\O')=H^0(\SIgma',\O)$, where $\SIgma'$ is a complex supermanifold
obtained from $\Sigma$ by blowing down the Ramond divisors  (this blowdown operation is discussed in \cite{WittenDonagi}, section
3.4.2).  The locally-free condition $H^0(\Sigma',\O)=\C$ is equivalent to the familiar 
condition   $H^1(\Sigma_\red,\R^{-1})=0$.  The last claim is shown as follows. It suffices to assume that $\Sigma$ is split, in which
case  $\Sigma'$ is the total space of the line
bundle $\varPi\R(p_1+\dots+p_{2r})\to \Sigma_\red$, so that an odd holomorphic function on $\Sigma'$ (which would obstruct the claim that
$H^0(\SIgma',\O)\cong \C$) corresponds to an element of $H^0(\Sigma_\red,\R^{-1}(-p_1-\dots-p_{2r}))$.  As a consequence of (\ref{zelmonk}),
we have \begin{equation}\label{welbo}H^0(\Sigma_\red,\R^{-1}(-p_1-\dots-p_{2r}))\cong H^0(\Sigma_\red,K\otimes \R).\end{equation} 
 By Serre duality, this vanishes if and only
if $H^1(\Sigma_\red,\R^{-1})=0$.  As long as this is true, the space of closed holomorphic 1-forms has the expected dimension..  

\subsection{Middle-Dimensionality Of The Periods}\label{nonperiods}

In defining the super period matrix in section \ref{superp}, we assumed that the space $\Lambda_0$ of periods of closed holomorphic 1-forms is
middle-dimensional in the space $\Lambda$ of periods. We will now show that this is true for any split super Riemann surface $\Sigma$.  As long as the
locally-free condition $H^0(\SIgma,\O')\cong \C$ is satisfied, $\Lambda_0$ automatically remains middle-dimensional when $\Sigma$ is deformed
away from the split locus.
(Since odd moduli are infinitesimal, turning them on  will not cause a nonzero period to become zero, so it will not reduce the dimension
of $\Lambda_0$.  On the other hand,
the bilinear relations (\ref{bilrel}) ensure that the dimension of $\Lambda_0$ cannot increase.)   For $\Sigma$ split,
the following analysis will show that $\Lambda_0$ is middle-dimensional even for  $H^1(\SIgma_\red,\R^{-1})\not=0$,
but in that case, we cannot say
anything simple about what happens away from the split locus.

Middle-dimensionality of the even periods on a split super Riemann surface $\Sigma$ just amounts to the classical fact that on an ordinary Riemann surface $\SIgma_\red$, the periods of holomorphic
differentials are middle-dimensional in the space of all $A$- and $B$-periods. We will now show that the same is true for the odd periods.

For a line bundle $\L\to \Sigma_\red$, we write $h^i(\L)$ for the dimension of $H^i(\Sigma_\red,\L)$.  Since $\R^{-1}$ has degree $g-1+r$,
the Riemann-Roch theorem gives $h^0(\R^{-1})-h^1(\R^{-1})=r$.  Via Serre duality, this is equivalent to $h^0(\R^{-1})-h^0(K\otimes \R)=r$. In
view of (\ref{welbo}), this is equivalent to
\begin{equation}\label{welmonk}h^0(\R^{-1})-h^0(\R^{-1}(-p_1-\dots-p_{2r}))=r. \end{equation}
A section of $\R^{-1}(-p_1-\dots -p_{2r})$ is simply a section of $\R^{-1}$ that vanishes at $p_1,\dots,p_{2r}$, so $H^0(\Sigma_\red,\R^{-1}
(-p_1-\dots -p_{2r}))$ is a subspace of $H^0(\Sigma_\red,\R^{-1})$.  Eqn. (\ref{welmonk}) says that the quotient space
has dimension $r$:
\begin{equation}\label{molfonk}\mathrm{dim}\,\left(H^0(\Sigma_\red,\R^{-1})/H^0(\SIgma_\red,\R^{-1}(-p_1-\dots-p_{2r}))\right)=r.\end{equation}
Now, $H^0(\Sigma_\red,\R^{-1})$ is the space of even closed holomorphic 1-forms, and $H^0(\Sigma_\red,\R^{-1}(-p_1-\dots -p_{2r}))$ is
its subspace consisting of those that vanish when restricted to Ramond divisors or in other words whose odd periods vanish.
So eqn. (\ref{molfonk}) says that for any split super Riemann surface $\Sigma$, the space of even closed holomorphic 1-forms modulo those
with vanishing periods is of dimension $r$. 

In other words, for $\Sigma$ split or for $H^0(\Sigma,\O')\cong \C$, the periods of closed holomorphic 1-forms always span a middle-dimensional
subspace  $\Lambda_0\subset \Lambda$.

\subsection{More On The Split Case}\label{anex}

Next we will look more closely at the super period matrix of a split super Riemann surface $\SIgma$. 
Odd closed holomorphic 1-forms on $\SIgma$ are simply pullbacks of holomorphic 1-forms on $\SIgma_\red$.
Their periods are just the corresponding periods on $\SIgma_\red$.  So the pseudoclassical block $\h\Omega_{g\times g}$ of the
period matrix of $\SIgma$ is just the classical period matrix of $\Sigma_\red$.  A form on $\Sigma$ that is a pullback from $\SIgma_\red$
has no $\d\theta$ component. So its odd periods vanish, and hence $\h\Omega_{g\times r}=\h\Omega_{r\times g}=0$.  Thus the
super period matrix of $\Sigma$ is
\begin{equation}\label{orbo}\h\Omega=\begin{pmatrix}\h\Omega_{g\times g}& 0 \cr 0 & \h\Omega_{r\times r}\end{pmatrix},\end{equation}
where only $\h\Omega_{r\times r}$ remains to be understood.

For this, we first recall that an even holomorphic 1-form on $\Sigma$ is exact, $\nu=\d(g(z)\theta)$ for
some $g(z)$, so its ordinary $A$- and $B$-periods -- that is, its even periods -- vanish. (This gives  another explanation of why $\h\Omega_{g\times r}=\h\Omega_{r\times g}=0$.)
Now suppose that
$\nu$ and $\nu'$ are two even closed holomorphic 1-forms on $\Sigma$, with respective 
odd periods $w_\alpha$ and $w'_\alpha$.  
We combine the odd periods of $\nu$ and $\nu'$  
 into vectors $\upnu, \,\upnu'$, which
take values in a vector space $\varLambda$
of dimension $2r$ that has a basis corresponding to the oriented Ramond divisors $\F_\alpha$.  Specialized to the case that
the even periods are zero, the bilinear relation of eqn. (\ref{bilrel}) reduces to
\begin{equation}\label{mondo}\sum_\alpha w_\alpha w'_\alpha=0. \end{equation}
We can see very directly why this is true.   Suppose that $\nu=\d(g\theta)$, $\nu'=\d(g'\theta)$, with $g,g'\in H^0(\Sigma_\red,\R^{-1})$.
The product $gg'$ is then a section of $\R^{-2}$, but the isomorphism in (\ref{elmonk}) identifies this
with $K\otimes \O(p_1+\dots+p_{2r})$.  A section of $K\otimes \O(p_1+\dots+p_{2r})$ is a meromorphic 1-form that may have simple
poles at the points $p_1,\dots,p_{2r}$; the residue of the pole of $gg'$ at $z=z_\alpha$ is the product $w_\alpha w'_\alpha/2\pi\sqrt{-1}$.
Thus eqn. (\ref{mondo}) asserts the vanishing of the sum of residues of a certain meromorphic 1-form.

For an even more explicit example, we consider a super Riemann surface 
$\Sigma$ of genus 0 with two Ramond punctures.
We parametrize $\Sigma$ -- or more precisely, the complement of a divisor in $\Sigma$ -- by coordinates $z|\theta$ 
with superconformal structure defined by
\begin{equation}\label{zeldo}\varpi^*=\d z-z\theta\d\theta.\end{equation}
The Ramond punctures are at $z=0$ and $z=\infty$.  To understand what is happening at $z=\infty$, we 
introduce new coordinates via $z=1/y$, $\theta=\sqrt{-1}\psi$,
whence 
\begin{equation}\label{weddo} \varpi^*=-\frac{1}{y^2}\left(\d y - y\psi\d\psi\right) .\end{equation}
The factor of $-1/y^2$ is not important here, since we only care about the subbundle of the cotangent bundle of 
$\Sigma$ that is generated by $\varpi^*$.
Thus the superconformal structure near $y=0$ is generated by $\d y-y\psi\d\psi$, showing that $y=0$ is a Ramond 
divisor and that the coordinate system
$y|\psi$ puts the superconformal structure in a standard form near this divisor.  Notice that the factor of $\sqrt{-1}$ in 
the formula $\theta=\sqrt{-1}\psi$ is necessary
for this result.  There are no nonzero odd closed holomorphic  1-forms on $\Sigma$, and the space of even ones is one-dimensional,
generated by $\nu=\d\theta=\sqrt{-1}\d\psi$.  The odd periods of $\nu$ are 1 at $z=0$ and $\sqrt{-1}$ at $y=0$, 
so the sum of squares of the odd periods
is zero, as expected.

Of course, we could also describe the same example more globally in projective coordinates.  For this, we simply take $\Sigma$
to be a weighted projective superspace $\WCP^{1|1}(1,1|0)$ with even homogeneous coordinates $u,v$ of weight 1 and an odd homogeneous
coordinate $\theta$ of weight 0.  The superconformal structure of $\Sigma$ can be defined by the following section of $T^*\SIgma\otimes \O(2)$:
\begin{equation}\label{definit}\h\varpi= u\d v - v\d u - uv \theta\d\theta. \end{equation}

More generally, we can describe a split genus 0 super Riemann surface $\Sigma$ with $2r$ Ramond punctures in affine coordinates $z|\theta$
by the superconformal structure
\begin{equation}\label{efinit}\varpi^*=\d z -\prod_{k=1}^{2r}(z-e_k)\theta\d\theta . \end{equation}
Alternatively,  $\SIgma$ is the weighted projective space $\WCP^{1|1}(1,1|1-r)$ with homogeneous
coordinates $u,v|\theta$ of weights $1,1|1-r$ and superconformal structure defined by $\h\omega=u\d v-v\d u -P(u,v)\theta\d\theta,$ with $P(u,v)$ a homogeneous
polynomial of degree $2r$. 

It is instructive to consider this last example more carefully to see what is involved in orienting the Ramond divisors.  To orient
the Ramond divisor $\F_k$ at, say, $z=e_k$, we should pick new coordinates $z|\h\theta_k$ such that $\varpi$
takes the standard form $\d z -(z-e_k)\h\theta_k\d\h\theta_k$ near $z=e_k$.  Then we orient $\F_k$ by choosing the differential
$\d\h\theta_k$ along $\F_k$.  Clearly, we need
\begin{equation}\label{melod}\h\theta_k=\theta\cdot \sqrt{\prod_{k'\not=k} (e_k-e_{k'})}.  \end{equation}

Now let us ask what happens to the orientations of the Ramond divisors $\F_k$  when the moduli $e_1,\dots,e_k$ are varied.  
To keep things simple, we consider the Ramond divisors to be labeled (as is usually most natural in string theory), so we do not allow permutations of the $e_k$; we consider
only what happens when the $e_k$ are braided around each other.  Such a process is made by composing elementary moves
 in which one of the $e$'s, say
$e_{k_1}$, makes a small loop around another, say $e_{k_2}$.  In the process, $\sqrt{e_{k_1}-e_{k_2}}$ changes sign, but there
are no sign changes in $\sqrt{e_i-e_j}$ for any other pair.  So the orientations of $\F_{k_1}$ and $\F_{k_2}$ are reversed, and no others.  
Combining any number of operations of this kind, we see that the only constraint is that there are an even number of orientation
reversals.  In other words, the only constraint is that the 
monodromies preserve the orientation of the space $\varLambda$ that parametrizes the odd periods.

This has a simple explanation.  $\varLambda$ is an even-dimensional vector space with a non-degenerate quadratic form.  In such
a vector space, there are two families of maximal isotropic subspaces, associated to a choice of orientation.  $\varLambda$ has a distinguished
middle-dimensional isotropic subspace, the space $\varLambda_0$ of odd periods of odd holomorphic differentials.  So it has a preferred
orientation that is preserved under any monodromies.

\subsection{Two Ramond Punctures}\label{tworp}

Our application in the remainder of this paper will involve the case of a super Riemann surface $\Sigma$ with precisely 2 Ramond
punctures.  So let us point out some particularly nice things that happen in this case.

With two Ramond punctures, there are precisely 2 odd periods, say $w_1$ and $w_2$, so the space $\varLambda$
of odd periods only has two null subspaces, generated by $w_1\pm \sqrt{-1}w_2$.  The choice of a fermionic $A$-period
in this case (eqn. (\ref{melz})) is particularly simple.  There is only one fermionic $A$-period $w$, and up to an integer power
of $\sqrt{-1}$ (which will arise if we exchange $w_1$ and $w_2$, or reverse their orientations), it must be
either $\frac{1}{\sqrt 2} (w_1+\sqrt{-1}w_2)$ or $\frac{1}{\sqrt 2}(w_1-\sqrt{-1}w_2)$.  

However, only one of the two choices is viable in the definition of the super period matrix.  According to eqn. (\ref{omorb}),
we are supposed to find an even closed holomorphic 1-form $\nu$ with $w(\nu)=1$.  However, for $\Sigma$ split, $\nu$ is a null vector in
$\varLambda$, so its periods obey $w_2=\pm \sqrt{-1}w_1$, with one choice of the sign or the other.  This means that if we choose
the wrong sign in the definition of the fermionic $A$-period $w$, then we will get $w(\nu)=0$ and will be unable to satisfy $w(\nu)=1$.

So a unique definition of $w(\nu)$ is forced upon us, up to an integer power of $\sqrt{-1}$.  Moreover, with this choice, as long
as the space of closed holomorphic 1-forms has the expected dimension $g!1$, a unique set of forms obeying (\ref{zorb}) and (\ref{omorb})
always exists,  Hence, the super period matrix is always defined away from the usual locus $\WW$ along which $H^0(\Sigma,K\otimes \R^{-1})
\not=0$.  

If we do multiply $w$ by $\sqrt{-1}^a$, for some integer $a$, what happens to $\h\Omega$?  To compensate for the change
in $w$, we will have to multiply $\t w$ and $\nu$ by $\sqrt{-1}\,^{-a}$.  $\h\Omega_{r\times r}$  is multiplied by $(-1)^a$
and $\h\Omega_{r\times g}$ and $\h\Omega_{g\times r}$  are multiplied by $\sqrt{-1}\,^{-a}$.  $\h\Omega_{g\times g}$  is unchanged.  

With more than 2 Ramond punctures, there are more choices in defining the fermionic $A$-periods.  The nondegeneracy condition
in the definition of the super period matrix is more complicated, and the super period matrix has poles when this condition fails.  We consider this next.

\subsection{More Than Two Ramond Punctures}\label{nond}

In the definition of the super period matrix, we needed to know that closed holomorphic 1-forms obeying the conditions (\ref{zorb})
and (\ref{omorb}) exist and are unique.  Saying that a system of linear equations (with the same number of variables and unknowns)
has a unique solution is an open condition, so it suffices to consider the case that $\Sigma$ is split.  Then the condition
is simply that it should be possible to find an even differential $\nu$ with prescribed values of half of its odd periods $w^\eta$, $\eta=1,\dots,r$ 
(and no condition on the other odd periods $\t w_\eta$).   (This is clearly equivalent to the existence of  differentials $\nu_\zeta$
with $w^\eta(\nu_\zeta)=\delta^\eta_\zeta$.)  On dimensional grounds, an equivalent statement is the following.  Let 
$\varLambda\cong \C^{2r}$ have a basis corresponding to oriented Ramond divisors, let 
$\varLambda_0$
be the middle-dimensional isotropic subspace of $\varLambda$ that parametrizes odd periods of closed holomorphic 1-forms, and let $\varLambda_1$ be the middle-dimensional isotropic subspace of $\varLambda$ characterized by $w^1=\dots=w^r=0$.  Then
the desired condition is $\varLambda_0\cap \varLambda_1=0$.  

As remarked at the end of section \ref{anex}, middle-dimensional isotropic subspaces of $\varLambda$ come in two families
associated with a choice of orientation of $\varLambda$.  A necessary condition for $\varLambda_0\cap\varLambda_1=0$ is
that $\varLambda_0$ and $\varLambda_1$ should be oppositely oriented (that is, associated to opposite orientations of $\varLambda$)
if $r$ is odd, or oriented the same way if $r$ is even.  Conversely, if the orientations of two middle-dimensional isotropic subspaces
$\varLambda_0$ and $\varLambda_1$ make this
possible, then generically $\varLambda_0\cap \varLambda_1=0$. 

One may therefore expect that as long as the right orientation is used in defining the fermionic $A$-periods, the nondegeneracy condition
$\varLambda_0\cap\varLambda_1=0$ will be satisfied generically, on the complement of a divisor in the reduced space of $\MM_{g,0,2r}$.
To show that this is true, it suffices to show it for $g=0$, since a super Riemann surface of any genus with $2r$ Ramond punctures
can degenerate to several components one of which is a  genus 0 surface containing all of the Ramond punctures.  Using the explicit
description  (\ref{efinit}) of a super Riemann surface of genus $g$ with $2r$ Ramond punctures, one can show directly that, with the right
definition of the fermionic $A$-periods, the nondegeneracy condition is obeyed generically.  Pick $r$ distinct points $f_1,\dots,f_r\in \C$
and consider a limit with $e_{2i-1},e_{2i}$ near $f_i$ for $i=1,\dots,r$.    Explicitly we find  that in this limit, with fermionic $A$-periods
$w^i$ defined as in
eqn. (\ref{melz}), the  differentials $\nu_k$ that satisfy $w^i(\nu_k)=\delta^i_k$ are $\nu_k=\d(a_k(z)\theta)$ with
\begin{equation}\label{exx}a_k(z)\sim \frac{\sqrt{e_{2k-1}-e_{2k}}}{\sqrt{4\pi \sqrt{-1}}}\prod_{j\not=k}(z-f_j).\end{equation}
We do not know a useful characterization of the divisor on which  $\varLambda_0\cap\varLambda_1\not=0$, producing additional
poles in the super period matrix.

\subsection{The Bad Set}\label{badset}

Here we will make some observations about the bad set $\WW$ in moduli space along which $h^0(\R^{-1})>r$.

Let  $\L\to \Sigma_\red $ be a line bundle of degree $g-1+s$, determined by a point in $\Jac_{g-1+s}$, the component of the Jacobian
that parametrizes line bundles of that degree.  Generically, the condition $h^0(\L)>s$ is satisfied only in codimension $s+1$ in $\Jac_{g-1+s}$.
For example, if $s=g-1$, then $h^0(\L)\geq s+1=g$ if and only if $\L\cong K$, which determines a unique point in $\Jac_{g-1+s}$, of codimension
$g$.  

However, we are interested in a line bundle $\R^{-1}$ with an isomorphism
\begin{equation}\label{monx}\R^{-2}\cong K(p_1+\dots+p_{2r}).  \end{equation}
This case is somewhat exceptional for small $r$.  We will examine this in detail, since our application later in this paper involves $r=1$.

For $r=0$, $\R^{-1}$ is simply a square root of $K$.  The codimension along which $h^0(\R^{-1})\not=0$ is 0 or 1 depending on
whether $\R^{-1}$ defines an odd or even spin structure.  

The case $r=1$ is somewhat similar to $r=0$ with odd spin structure: the condition $h^0(\R^{-1})\geq r+1=2$ is satisfied in codimension
1, not in the ``expected'' codimension $r+1=2$.  Before explaining this in detail, we first consider some small genus cases that are relevant to our
applications.

Our main application later in this paper involves genus 2.
A genus 2 Riemann surface $\Sigma_\red$ is hyperellipic, and is a two-fold cover $\rho:\Sigma\to \CP^1$.   For $r=1$, the line bundle
$\R^{-1}$ has degree 2. The Riemann-Roch formula gives $h^0(\R^{-1})-h^1(\R^{-1})=1-g+\mathrm{deg}\,\R^{-1}=1$, and Serre duality gives $h^1(\R^{-1})=h^0(K\otimes \R)$.  So  $h^0(\R^{-1})\geq 2$  is equivalent to $h^0(K\otimes \R)>0$.
But $K\otimes \R$ is of degree 0.  A line bundle of degree 0 with a holomorphic section must be trivial, so $K\otimes \R\cong \O$ and
$\R^{-1}\cong K$.  Since also $\R^{-2}\cong K(p_1+p_2)$, we must have $K\cong \O(p_1+p_2)$.  As explained in section \ref{specialcase}, 
this is so if and only if the two points $p_1,p_2$
are exchanged by the hyperelliptic involution of $\SIgma$.   There is a one-parameter family of such pairs.  This family is of codimension 1
in the space of all pairs $p_1,p_2$, so the exceptional set $\WW$ is a divisor $\DD$ in this case.

We will also consider in section \ref{genthree} the case of genus 3.  A generic genus 3 Riemann surface $\Sigma_\red$, 
in affine coordinates, is described as a plane curve $P_4(x,y)=0$, where $P_4$ is a quartic polynomial in two
variables.  A canonical divisor is the intersection of $\Sigma_\red$ with a line $L$ in the plane. A line is of course defined by a linear
equation $P_1(x,y)=0$.   Pick any point $w\in\Sigma_\red$
and let $L$ be the line tangent to $\Sigma_\red$ at $w$.   The equations $P_1(x,y)=P_4(x,y)=0$, which describe intersections of $\Sigma$ and $L$,
 will be satisfied at four points in $\C^2$,
counted with multiplicity.  The point  $w$ of tangency has multiplicity 2, so $\Sigma_\red$ and $L$ intersect at two other points $p_1,p_2$.  Generically
these are distinct points with multiplicity 1 each.  So $K\cong \O(2w+p_1+p_2)$.  Thus the line bundle $\R^{-1}=\O(w+p_1+p_2)$ admits
an isomorphism $\R^{-2}\cong K(p_1+p_2)$.  The line bundle $\R^{-1}(-p_1-p_2)\cong \O(w)$ has a non-zero holomorphic section (the section ``1''  that
vanishes precisely at $w$).  Since $\O(w)$ has degree 1, we have  $h^0(\O(w))=1$ (a  degree 1 line bundle $\L$ over a curve of positive genus 
always has $h^0(\L)\leq 1$) and the Riemann-Roch formula $h^0(\O(w))-h^1(\O(w))=1-g+1=-1$ implies that  $h^1(\O(w))=2$.  But by Serre duality
$h^1(\O(w))=h^0(K(-w))=h^0(\R^{-1})$.  So $h^0(\R^{-1})=2$.  Thus  we have found a 1-parameter family of pairs $p_1,p_2$, parametrized by
$w\in\Sigma$, such that $h^0(\R^{-1})\geq r+1=2$, showing that again the exceptional set $\WW$ is of codimension 1.

Now let $\Sigma_\red$ have arbitrary genus $g$.  For $r=1$, the line bundle $\R^{-1}(-p_1)$ has degree $g-1$, so one expects that there is a divisor $\DD$ in the moduli space along which 
$h^0(\R^{-1}(-p_1))>0$.  Along $\DD$, let $s$ be a nonzero holomorphic section of $\R^{-1}(-p_1)$.  Then $s^2$ is a holomorphic section of
$\R^{-2}(-2p_1)\cong K(-p_1+p_2)$.  In other words, $s^2$ is a meromorphic section of $K$ that is holomorphic except possibly for a single pole at $p_2$.
Since the sum of residues of a meromorphic section of $K$ must vanish, $s^2$ is actually holomorphic at $p_2$.  This means that $s$ must vanish
at $p_2$, so $s$ is a holomorphic section of $\R^{-1}(-p_1-p_2)$.  Thus along $\DD$, $h^0(\R^{-1}(-p_1-p_2))>0$. Since $\R^{-1}(-p_1-p_2)$ has
degree $g-2$, Riemann-Roch implies that $h^0(\R^{-1}(-p_1-p_2))-h^1(\R^{-1}(-p_1-p_2))=-1$, so $h^1(\R^{-1}(-p_1-p_2))\geq 2$.  
By Serre duality, this is equivalent to $h^0(K\otimes \R(p_1+p_2))\geq 2$.  Finally, using the isomorphism (\ref{monx}), this is equivalent to
$h^0(\R^{-1})\geq 2$.  In short, along the divisor $\DD$, one has $h^0(\R^{-1})\geq 2$.  Along this divisor, the super period matrix has a pole with
nilpotent residue, as will be described in Appendix \ref{perturb}.

For $r>1$, it is likely that the exceptional set $\WW$ has codimension greater than 1, but we will not analyze this case in detail.  We should
note that although the choice of fermionic $A$-periods is essentially unique for $r=1$, as described in section \ref{tworp}, for $r>1$,
one requires a somewhat arbitrary choice of fermionic $A$-periods and this introduces   artificial singularities in codimension 1.  Thus describing
the periods by a period matrix is less natural for $r>1$ than it is for $r=1$.   It is perhaps more natural for $r>1$ to simply study the Lagrangian
submanifold spanned by the periods, rather than to define a period matrix.

\section{Low Genus}\label{twop}

In the rest of this paper, we navigate toward an application of the super period matrix with Ramond punctures that was described in the
introduction.  The application mainly involves super Riemann surfaces of genus 2, so we begin by explaining some special facts about
the super period matrix for small values of the genus.  

 In genus $g\leq 3$, any $g\times g$ complex symmetric matrix  with positive imaginary part is the period matrix of an
ordinary Riemann surface $\Sigma_0$, which is unique up to isomorphism.  (This is not true for $g>3$; 
for a symmetric matrix of positive imaginary part to be a period matrix, it must obey
the Schottky relations. That
is why the following construction is limited to $g\leq 3$.)  
If $\Sigma_0$ is a Riemann surface of genus $\leq 2$ with even spin structure, then $H^0(\Sigma_0,K^{1/2})$ 
is zero always\footnote{\label{zelf} This statement is true for a smooth Riemann surface of genus 2.  However, a smooth curve of genus
2 with even spin structure can degenerate to a pair of genus 1 components each with odd spin structure, meeting at a point; for such a singular curve, the appropriate
analog of $H^0(\Sigma_0,K^{1/2})$ is non-zero.  Thus the divisor $\DD$ has a component at infinity in the Deligne-Mumford
compactification of $\M_{2,\spin +}$. } (and not just generically,
as is the case for $g\geq 3$), so any super Riemann surface of genus 2 with even spin structure has a super period matrix $\h\Omega$.
By mapping a super Riemann surface $\Sigma$ to the ordinary Riemann surface with the same period matrix, we get for $g=1,2$
a natural holomorphic map\footnote{For $g\leq 3$, we generically use the symbol $\pi$ to denote a projection defined using the period matrix,
or its pseudoclassical block in the presence of Ramond punctures.} $\pi:\MM_{g,+}\to \M_{g,\spin +}$, where $\MM_{g,+}$ parametrizes super 
Riemann surfaces of genus $g$ with even
spin structure, and its reduced space $\M_{g,\spin+}$ parametrizes an ordinary Riemann surface of genus $g$ also with even spin structure.

For $g=1$, this construction is trivial, as a genus 1 super Riemann surface with even spin structure 
has no odd moduli and $\MM_{1,+}=\M_{1,\spin+}$.
However, for $g=2$, the statement is non-trivial and
has been exploited by D'Hoker and Phong in computing superstring scattering
amplitudes, as summarized in \cite{DPh}.  We will ultimately focus mostly on this case.  For $g=3$, since every $3\times 3$ complex 
matrix of positive definite imaginary part is a period
matrix, we can use the same construction to define a meromorphic projection $\pi:\MM_{3,+}\to \M_{3,\spin+}$, 
but now $\pi$ has poles (with nilpotent residue).   As explained in Appendix \ref{hyperelliptic}, 
 $\pi$ has for genus 3 a fairly obvious pole along the divisor $\DD$ where the super period matrix has a pole,
and a somewhat less obvious pole along a second divisor $\DD'$.

We can  do something somewhat similar for $\MM_{g,0,2r}$, which parametrizes a genus $g$ super 
Riemann surface $\Sigma$ with $2r$ Ramond punctures
and no NS punctures.  
Once we pick a set of fermionic $A$-periods, we can define a  super period matrix $\h\Omega$, 
which in particular has the $g\times g$ pseudoclassical
 block $\h\Omega_{g\times g}$.    For $g\leq 3$, 
mapping $\Sigma$ to an ordinary Riemann surface $\Sigma_0$ whose period matrix $\Omega$ coincides with $\h\Omega_{g\times g}$ gives a map
\begin{equation}\label{moxo}\begin{matrix}\X & \to & \MM_{g,0,2r}\cr
                                                                       && \downarrow\pi \cr
                                                                        && \M_g. \end{matrix}\end{equation}
The fiber $\X$ parametrizes all moduli of $\SIgma$ other than its super period matrix. 
Note that in 
(\ref{moxo}), the base space is simply $\M_g$, with no
memory of the generalized spin structure.  In the absence of Ramond punctures, the analogous projection $\pi:\MM_{g,\pm}\to \M_{g,\spin\pm}$ can
be defined to remember the spin structure.  We cannot do something analogous in the case of a projection from $\MM_{g,0,2r}$ to
$\M_g$; since the definition
of a generalized spin structure (a line bundle $\R$ with the isomorphism in eqn. (\ref{elmonk}))
depends on the positions of the Ramond punctures, there is no way to  forget
the Ramond punctures while remembering the generalized spin structure.  That is why anything that one can deduce from the fibration $\pi$ 
for $r>0$ -- notably the vanishing under certain conditions of the dilaton tadpole -- will involve a sum over spin structures.

The nicest case of the fibration $\pi:\MM_{g,0,2r}\to \M_g$ -- and the case that we will use in our application -- is for $r=1$, for then the choice of a fermionic
$A$-period is essentially unique (and $\h\Omega_{g\times g}$ is entirely unique) as we saw in section \ref{tworp}.  For $r>1$, the projection 
$\pi$ does depend on the choice of fermionic $A$-periods and moreover it has unphysical singularities that depend on that choice.

Keeping $g\leq 3$, there is no trouble to include NS punctures in this picture, since an 
NS puncture is just a marked point.  There is therefore a forgetful map for NS punctures, and an NS puncture
does not affect the definition of the super period matrix. 
Composing the map $\MM_{g,n,2r}\to \MM_{g,0,2r}$  which forgets
the NS punctures with the  projection            $\pi$,        we get a projection                                       
 \begin{equation}\label{moxop}\begin{matrix}\Y & \to & \MM_{g,n,2r}\cr
                                                                       && \downarrow\pi' \cr
                                                                        && \M_g, \end{matrix}\end{equation}   
where now the fiber $\Y$ parametrizes also the positions of the NS punctures.  

Our application will involve the case $r=1$ (precisely 2 Ramond punctures) and $n=0$ (no NS punctures).  We will also  encounter
the case $r=0$, $n=1$. The projection $\pi:\MM_{2,0,2}\to\M_2$  that we will use has poles
 associated to poles of the super period matrix.
The complications associated to those poles can be overcome, at least for certain purposes.
We will analyze this in section \ref{lowgenus},
but there are a number of things to explain first.

All statements in this section have been made for the uncompactified moduli spaces. Some care is required to extend these 
statements over the corresponding Deligne-Mumford compactifications.  
For example, in the absence of Ramond punctures, the projection $\MM_{2,+}\to \M_{2,\spin+}$ used
by d'Hoker and Phong develops a pole if one attempts to extend it over the Deligne-Mumford compactifications of 
these spaces.  This has been explained in   \cite{Wittenmore} (the underlying reason was explained in footnote \ref{zelf} above),
and is important in understanding the behavior near infinity of the measure on $\M_{2,\spin+}$ computed by d'Hoker and Phong.

\section{Ward Identities}\label{app}

A certain subtlety of superstring theory is important in our application.  
Before explaining it, we begin by recalling what happens in bosonic string theory.

Consider a closed oriented bosonic string vacuum
such that at  tree level the matter system has a  continuous symmetry,\footnote{$D$-branes and/or an orientifold projection can be included
in the following discussion if 
the $D$-brane boundary condition and/or the orientifold projection are chosen to preserve the symmetry.} with a conserved charge $\J$ associated to a  
conserved worldsheet current $J$.  

Such a symmetry constrains the worldsheet correlation functions on any worldsheet.  Let $V_1,\dots,V_n$ be an arbitrary set of matter vertex
operators that are eigenstates of $\J$ (and are conformal primaries of the appropriate dimension).  
Then the correlation function $\langle V_1V_2\dots V_n\rangle$ on a worldsheet $\SIgma_0$ of any genus $g$ will
vanish unless the sum of the charges of the operators $V_i$ vanishes.  Thus, the existence of a conserved 
worldsheet current leads automatically to a conservation
law for genus $g$ amplitudes, for any $g$.  As a result, in closed oriented bosonic string theory, a continuous symmetry 
at tree level remains valid
as a continuous symmetry to all orders of perturbation theory.  This result holds for any given Riemann surface $\Sigma_0$; it does not
in any way involve integration over moduli space.  

 The same argument holds for continuous symmetries 
of superstring theory that arise from
the NS sector.\footnote{If (as is typical) the continuous symmetry is associated at tree level to a massless gauge particle, the gauge field may gain mass in perturbation theory
but the symmetry is unbroken as a global symmetry to all orders of perturbation theory.  
See \cite{Wittenmore} for further detail.}  Thus, loop corrections do not trigger spontaneous breakdown of a continuous global symmetry that
arises in closed oriented bosonic string theory or the NS sector of superstring theory.

However, we cannot make such a simple argument for spacetime supersymmetry, which comes from the Ramond sector.  The analog of the conserved current $J$ is the usual fermionic
vertex operator $S_A$ (here $A$ is a spinor index and $S_A$ is the combined spin field of matter and ghosts).  The considerations
of this paper apply equally to heterotic or Type II superstrings, but for simplicity, and also
with a view toward our eventual application, we consider the heterotic string,
in which spacetime supersymmetry comes from the holomorphic part of the worldsheet theory only.  Then $S_A$ is a holomorphic object, but it
cannot be understood as a conserved current on a fixed super Riemann surface $\Sigma$; indeed,
 it is inserted at a Ramond puncture, which is a 
singularity in the superconformal structure of $\Sigma$, and there is no notion of moving a Ramond puncture while otherwise leaving $\Sigma$ unchanged.

Instead, the proof of a supersymmetric Ward identity proceeds essentially by constructing a conserved current on the super moduli space rather than on the super
Riemann surface $\Sigma$.  To do this, we combine the spin field with the ordinary $c$ ghost to make 
a dimension $(0,0)$ superconformal primary $\S_A=c S_A$.  We similarly
combine the matter vertex operators $V_i$ with ghosts in the usual way to make superconformal primaries 
$\V_i$ of dimension $(0,0)$.  (For example, if $V_i$
is a vertex operator of the NS sector, then we set $\V_i=\t c c\delta(\gamma)V_i$, where  $c,\gamma$ are the 
holomorphic superconformal ghost fields
and $\t c$ is the antiholomorphic ghost field.)  The correlation function 
\begin{equation}\label{morevert}F_{\S_A\V_1\dots\V_n}=\langle \S_A\V_1\dots\V_n\rangle \end{equation}
does not define a measure on supermoduli space -- or more precisely\footnote{As explained
in \cite{Wittenmore} and more fully in section 5 of \cite{Wittennotes} (see also \cite{DM}, pp. 94-5), 
the usual notion of integration over supermoduli space is an approximation to a 
more precise notion of integration over a certain cycle
$\varGamma$ in the product of holomorphic and antiholomorphic moduli spaces (in the case of the heterotic string, the 
holomorphic moduli space is the super moduli space
$\MM_{g,n,2r}$ and the antiholomorphic space is its reduced space with complex structure reversed; one can take
the reduced space of $\varGamma$ to coincide with the reduced space of $\MM_{g,n,2r}$).
The distinction between $\MM_{g,n,2r}$ and $\varGamma$ will  not be very important in the present paper, but we express our statements in terms of integration over $\varGamma$ since this is more accurate.} on the appropriate integration cycle $\varGamma$ of 
superstring perturbation theory -- because the operator
$\S_A$ has ghost number less by 1 than the ghost number of a standard Ramond-sector vertex operator.  As a result, $F_{\S_A\V_1\dots\V_n}$
represents not a measure on $\varGamma$ that could be integrated to compute a scattering amplitude, 
but an integral form of codimension 1 -- the supermanifold analog of a conserved current (see for example \cite{DM} or \cite{Wittennotes}).
BRST-invariance of $\S_A$ and of the $\V_i$ implies that this form is closed, $\d F_{\S_A\V_1\dots\V_n}=0$.  Using this fact and the supermanifold version
of Stokes's theorem, we derive a supersymmetric Ward identity:
\begin{equation}\label{melf}0=\int_{\varGamma}\d F_{\S_A\V_1\dots\V_n}=\int_{\partial\varGamma}F_{\S_A\V_1\dots\V_n}. \end{equation}
On the right hand side, $\partial\varGamma$ is a sum of components ``at infinity'' in $\varGamma$; they correspond to different ways that $\Sigma$ might
degenerate.

\begin{figure}
 \begin{center}
   \includegraphics[width=2.5in]{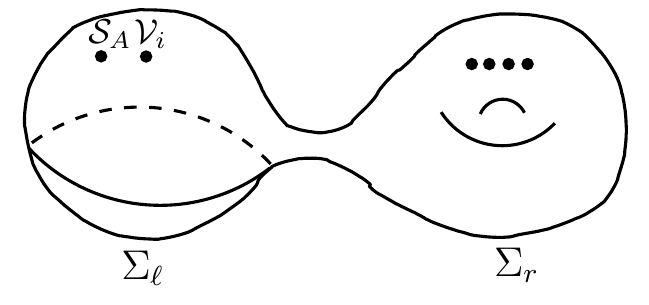}
 \end{center}
\caption{\small The Ward identity always receives contributions from separating degenerations of this kind in which  one
component $\Sigma_\ell$ contains a supersymmetry generator $\S$ and precisely
one more vertex operator. If these are the only contributions, then the Ward identity expresses the invariance of the $S$-matrix under spacetime supersymmetry.  The usual case, as sketched here, is that
$\Sigma_\ell$ has genus 0.  This leads to the familiar tree-level expressions for the supercharges.  }
 \label{Supersymmetry}
\end{figure}

\begin{figure}
 \begin{center}
   \includegraphics[width=2.5in]{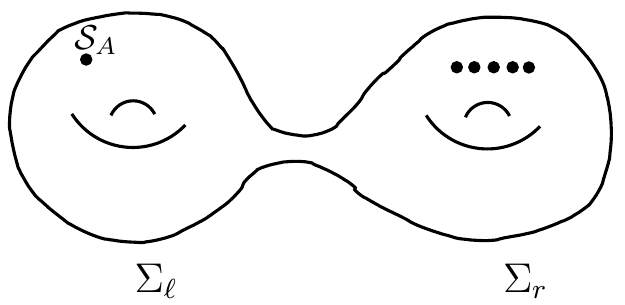}
 \end{center}
\caption{\small One other degeneration may contribute to the Ward identity.  This is
the Goldstone fermion contribution.  It represents spontaneous breaking of spacetime supersymmetry.
This contribution can exist only when the genus of $\Sigma_\ell$ is positive.}
 \label{Goldstone}
\end{figure}

Although in general $\Sigma$ has various separating and nonseparating degenerations, there are, as explained in \cite{Wittenmore} (and
more fully in section 8 of \cite{Revisited}), only two types of
degeneration that contribute to the Ward identity.  One such degeneration is the case that $\Sigma$ degenerates to
a union of two components, one of which contains $\S_A$ and just one of the $\V_i$.
Such a degeneration is  sketched in fig. \ref{Supersymmetry}.
As is explained in \cite{Wittenmore}, if only such components of  $\partial\varGamma$ contribute  on the right hand side of (\ref{melf}),
 then (\ref{melf}) becomes a standard Ward identity of unbroken supersymmetry and the genus $g$ contribution to the $S$-matrix is spacetime supersymmetric.  
 In general, there may also be a Goldstone fermion contribution to the Ward
identity; it can arise from a degeneration in which $\Sigma$ splits off a positive genus contribution $\Sigma_\ell$ that contains $\S_A$ but no other vertex
operator, as sketched in fig. \ref{Goldstone}.  Just as in field theory, if the genus $g$ Ward identity receives a Goldstone fermion contribution,
then the genus $g$ contribution to the $S$-matrix is not spacetime supersymmetric.  

The importance of the fibrations that were described in section \ref{twop} is that in favorable cases, 
they can be used to establish supersymmetric Ward identities in which one knows {\it a priori} that there can 
be no Goldstone fermion contribution.
  Rather than try to explain abstractly how that can happen, we will first describe the problem
that we have in mind.

\section{The Two-Loop Vacuum Amplitude}\label{twol}

We  turn to  the two-loop vacuum amplitude of heterotic string theory. Potential contributions  come from worldsheets with 
even spin structure only (worldsheets with odd spin structure contribute
to parity-violating amplitudes), so the vacuum amplitude is found by integration over $\MM_{2,+}$.  
An effective procedure \cite{DPh,DPhlatest}  has been to first integrate over the fibers of the projection 
$\pi:\MM_{2,+}\to \M_{2,\spin+}$, so as to reduce the vacuum
amplitude to an integral over the bosonic moduli space
$\M_{2,\spin+}$.  At that point, it makes sense to sum over spin structures
without integrating over any additional moduli. (This does not make sense before reducing to the bosonic 
moduli space, since there is no notion of changing the spin
structure of a super Riemann surface while otherwise leaving the surface unchanged.)
Quite a few supersymmetric compactifications to $\geq 4$ dimensions have been studied this way.   In each case, after reducing to 
$\M_{2,\spin+}$, the two-loop vacuum amplitude
vanishes upon summing over spin structures, even without any integral over the bosonic moduli.  
Our goal here is to explain this and show that it is true in general.

As we recalled in the introduction, in general, in a supersymmetric compactification to four dimensions, the procedure of integrating
over $\M_{2,\spin+}$ by integrating first over the fibers of the projection $\pi:\MM_{2,+}\to \M_{2,+}$ misses a contribution at infinity.
So our result means that in such compactifications, the full two-loop vacuum amplitude is given by the contribution at infinity.
In supersymmetric compactifications above four dimensions, there is no such contribution at infinity, and the whole two-loop vacuum
amplitude vanishes.

Before discussing the two-loop case, let us recall one way to understand what happens in genus 1.
On a superstring worldsheet $\Sigma$ of genus 1, we consider a two point function 
\begin{equation}\label{dolor}F_{\S_A\V^B}=\langle \S_A(z) \V^B(\t y;\neg y)\rangle. \end{equation}
Here as before $\S_A=cS_A$ is the spacetime supersymmetry generator with the ghost field $c$ included.  
  On the other
hand,  we take $\V^B$ to be the vertex operator for the dilatino -- the spin 1/2 partner of the dilaton -- at zero momentum.  For superstring theory in $\RR^{10}$, 
\begin{equation}\label{olor} \V^B=\t c c \delta(\gamma) \partial_{\t z}X^I\Gamma_I^{BC}S_C,\end{equation}
where $X^I,~I=1,\dots,10$ are matter superfields representing the motion of the strings in $\RR^{10}$,  $S_C$ is once again the holomorphic spin field, and $\Gamma_I^{BC}$, $I=1,\dots,10$, $B,C=1,\dots 16$ are gamma matrices.  In compactified 
models, the definition of $\V^B$ is changed slightly, but the details will not be important.    The notation $\S_A(z)$ and $\V^B(\t y;\neg y)$ is just meant to remind us that $\S_A$ is holomorphic
while $\V^B$ is neither holomorphic nor antiholomorphic.  

 These two vertex operators are both in the Ramond sector, so $\Sigma$ is a genus
1 surface with 2 Ramond punctures.  Such a surface has precisely 1 odd modulus.    The characteristic 
subtleties of superstring perturbation theory
result from the possibility of changes of variables such as $m\to m+\eta\eta'$, where $m$ is an even modulus and $\eta,\eta'$ are odd moduli.
So they do not come into play until there are at least two odd moduli.  Accordingly, none of these 
subtleties are relevant to the example under discussion.
Except for one important detail, we 
can think of $z$ and $y$ as points on the ordinary genus 1 Riemann surface $\Sigma_\red$, and 
use the translation symmetries of $\Sigma_\red$ to set $y=\t y=0$.
The detail in question is as follows: the generalized spin structure
of $\Sigma_\red$ changes when $z$  moves around a non-contractible loop in $\Sigma_\red$, and hence we should think of $z$ as a point in
a $2^{2g}=4$-fold unramified cover of $\Sigma_\red$, which we will call $\Sigma'_\red$, that 
parametrizes a point in $\Sigma_\red$ together with a generalized spin structure.
  The condition $z=0$ defines a single point on $\Sigma_\red$, but on $\Sigma'_\red$ it corresponds
to four possible points $p_1,\dots,p_4$, which are labeled by the four possible spin structures on $\Sigma$ or equivalently on $\Sigma_\red$.  
(The generalized spin structure of $\Sigma$
reduces to an ordinary spin structure for $z\to 0$, where the two Ramond punctures coincide.)  
The correlation function $\langle S_A(z)\,\V^B(0;\neg 0)\rangle$ is a holomorphic 1-form
on $\Sigma'_\red$  that
has poles only at the points\footnote{The residue of the pole at the point corresponding 
to the odd spin structure on $\Sigma$ vanishes because of
fermion zero modes in the matter system,
so there are really only three poles, not four.}  $p_i$.
  Because of the operator product relation
\begin{equation}\label{mondox}S_A(z)\V^B(0;\neg 0)\sim\frac{\delta^B_A}{z}\V_{\mathrm{Dil}}(0;\neg 0), \end{equation}
where $\V_{\mathrm{Dil}}$ is the dilaton vertex operator at zero momentum, the residue of each pole is the dilaton tadpole 
corresponding to the given spin structure.
The vanishing of the sum of the residues of the holomorphic 1-form $\langle S_A(z)\,\V^B(0;\neg 0)\rangle$ on $\Sigma'_\red$ 
means that the dilaton tadpole vanishes, after summing over spin structures
and before integration over bosonic moduli.  

A standard argument shows that in genus $g$, the dilaton tadpole is $2g$ times the vacuum amplitude.  So the genus 1 vacuum amplitude likewise
vanishes after summing over spin structures, but without integration over bosonic moduli.

The same argument does not work in genus $g>1$, because there is more than 1 odd modulus, 
and the subtleties of superstring perturbation theory do come into
play.  We cannot think of the fermion vertex operator $S_A$ as a conserved current on a fixed worldsheet; instead, as was summarized
in section \ref{app}, a proper general argument proceeds by applying the identity (\ref{melf}) to the 
correlation function $F_{\S_A\V^B}$ and analyzing all possible boundary
contributions, including contributions that involve degenerations of $\Sigma$.  
By such reasoning one can show that (in a supersymmetric compactification with no Goldstone fermion contribution),
 the genus $g$ vacuum amplitude vanishes after integration over all moduli.
There is in general no simpler version of this statement that involves integrating or summing over only some of the moduli.

However, for $g=2$ we can in fact imitate the classical genus 1 argument, with some more care, 
using the map $\pi:\MM_{2,0,2}\to \M_2$ described in
eqn. (\ref{moxo}).  Of course, we will have to take into account the fact that this map is only generically defined, 
but this turns out to cause no problem.  The obvious singularities of the correlation function $F_{\S_A\,\V^B}=\langle S_A(z)\,\V^B(0;\neg 0)\rangle$ arise
from a collision of the two vertex operators; such a collision gives a pole whose residue is the amplitude on $\MM_{2,1,+}$
associated to the dilaton tadpole.   We will reduce
from $\MM_{2,0,2}$ to $\MM_{2,1,+}$ by extracting this residue.  By adapting  the classical
argument that was explained above in genus 1, 
we will show that if one follows the procedure of d'Hoker and Phong to compute genus 2 amplitudes,
then the bulk contribution
to the dilaton tadpole -- and hence to the vacuum amplitude -- vanishes after integrating over the fiber of $\pi':\MM_{2,1,0}\to\M_2$, 
even before integrating over $\M_2$.  Concretely, integrating over the fiber of $\pi'$ means integrating over the position at which the
dilaton vertex operator is inserted, and summing over spin structures, but not integrating over the moduli of the underlying genus 2 surface
$\Sigma_\red$.  The bulk contribution to the genus 2 vacuum amplitude, computed with the  d'Hoker-Phong procedure,
vanishes before that last integration.  

To make the analysis, we write $\varXi$ for a generic fiber of the projection $\pi:\MM_{2,0,2}\to\M_2$, and we 
look at the behavior of $F_{\S_A\V^B}$ near $\varXi$.
We recall that, as discussed in section \ref{app}, $F_{\S_A\,\V^B}$ is not a volume form on $\MM_{2,0,2}$ 
but a form of codimension 1 (technically an integral form of codimension 1).   Near $\varXi$, we can factorize $F_{\S_A\V^B}$
as $\pi^*(\mu)\cdot  F^*_{\S_A\V^B}$, where $\mu$ is a volume form on $\M_2$ and $F^*_{\S_A\V^B}$ is a 
form of codimension 1 along the fibers
of $\pi$.   The reason that this factorization exists is that the missing ``index'' of the codimension 1 form $F_{\S_A\V^B}$ is 
tangent to the fibers of $\pi$,
not the base.  This index is missing because  $\S_A$ is missing a factor of the antiholomorphic ghost field $\t c$.  From an antiholomorphic
point of view, the map $\pi$ is just the forgetful map $\M_{2,2}\to \M_2$, and $\t c$ represents a 1-form dual to the 
motion of the operator $\S_A$ along the
fiber of this map.  So the missing index is tangent to the ``fiber'' of the fibration $\pi$, not to the base, and that is why $F_{\S_A\V^B}$ can be factored 
as the product of the pullback $\pi^*(\mu)$ of a full volume form $\mu$ on the base times a form $F^*_{\S_A\V^B}$ of 
codimension 1 in the fiber direction.
This factorization is not completely canonical, since $\mu$ could be multiplied by a nonzero function on the base of the fibration, but once
we restrict to a particular fiber $\varXi$,
  $F^*_{\S_A\V^B}$ depends on the choice of $\mu$ only
by an overall multiplicative constant, which will not affect what follows.

We can now make the same argument as in (\ref{melf}), but with the fiber $\varXi$ replacing the full integration cycle $\varGamma$, and $F^*_{\S_A\V^B}$
replacing $F_{\S_A\V^B}$.  Since $\d F^*_{\S_A\V^B}=0$, we have
\begin{equation}\label{mono}0=\int_{\varXi'}\d F^*_{\S_A\V^B} =\int_{\partial\varXi'} F^*_{\S_A\V^B}. \end{equation}
Here $\varXi'$ is defined by throwing away a small neighborhood of the singularities of $F^*_{\S_A\V^B}$ in $\varXi$, and
$\partial\varXi'$ is a union of the boundaries of those small neighborhoods.

\begin{figure}
 \begin{center}
   \includegraphics[width=3.5in]{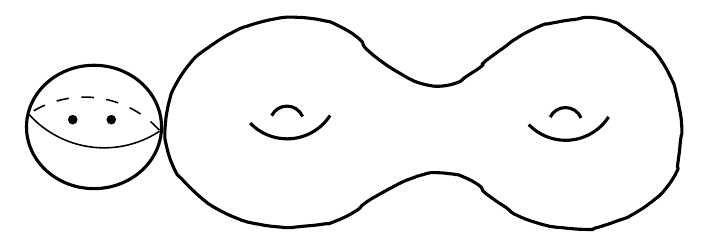}
 \end{center}
\caption{\small The obvious contributions to the identity (\ref{mono}) come from collisions of the two vertex operators.  From a conformal point of
view, these involve the splitting of $\Sigma$ into two components, of which one is a genus 0 surface with 2 Ramond punctures, one NS punctures,
and no even or odd moduli, and the second is a genus 2 surface with 1 NS puncture. The contribution of such a splitting to the identity
is the bulk contribution to the genus 2 dilaton tadpole, computed with the procedure of d'Hoker and Phong.  (The dilaton vertex operator is inserted on the genus 2
component at the point where it meets the second component.)} \label{splitting}
\end{figure}
The obvious singularities of $F^*_{\S_A \V^B}$ come from a collision of the two vertex operators $\S_A$ and $\V^B$.  Via the operator product (\ref{mondox}),
such a collision gives a pole whose residue is a dilaton tadpole evaluated on the surface $\Sigma$, endowed with one of its possible spin structures.
All possible spin structures occur because the fiber $\varXi$ parametrizes among other things the generalized
spin structure of the super Riemann surface $\Sigma$ (just as in genus 1,
the possible choices of this generalized spin structure are permuted when $\S_A$ or $\V^B$ is taken
around a noncontractible loop in $\SIgma$, and the generalized spin structure reduces to an ordinary one when the two operators meet).
 The locus in $\MM_{2,0,2}$ along which the two
vertex operators meet is  a divisor  $\frak D$ that is a copy of $\MM_{2,1,0}$ (fig. \ref{splitting}).  The projection $\pi:\MM_{2,0,2}\to \M_2$ which we use
to formulate eqn. (\ref{mono}) restricts along $\frak D$ to the projection $\pi':\MM_{2,1,0}\to \M_2$ that d'Hoker and Phong would use to compute the
bulk contribution to the dilaton tadpole.  So if the singularities that correspond to dilaton tadpoles are the only ones that contribute in the identity (\ref{mono}),
then we learn that with the d'Hoker-Phong procedure, the sum over spin structures of the genus 2 dilaton tadpole vanishes.  (The full answer
therefore comes from the correction at infinity that is described in \cite{Wittenmore}.)

It remains to consider the possibility of additional singularities contributing to the identity (\ref{mono}), resulting from the fact that the super period matrix
is only generically defined.  We will show in section \ref{lowgenus} that although there is indeed a locus on $\varXi$ on which the super period
matrix has a pole (with nilpotent residue), this leads to no contribution in the identity (\ref{mono}) because the fermions of the matter system acquire
zero-modes on just the dangerous locus.

 \section{What Happens At Poles Of The Super Period Matrix?}\label{lowgenus}                                                                       
 
In this section, we complete the argument of section \ref{twol}, by analyzing the singularities associated to poles of the super period
matrix, and showing that they do not affect the vanishing of the bulk contribution to the two-loop vacuum
amplitude.

 \subsection{The Locus Of Spurious Singularities}\label{specialcase}                                                                    
                                                  
As described in section \ref{badset}, in the case of a genus 2 surface $\Sigma$ with Ramond
punctures $p_1,p_2$, the condition $h^0(\R^{-1})>1$ is equivalent to $\O(p_1+p_2)\cong K$.  Moreover, this is
 also equivalent to $\O(p_1+p_2)\cong \R^{-1}$.
To make this condition more explicit, we recall that a genus 2 Riemann surface $\Sigma_\red$ is hyperelliptic; it admits a 
holomorphic map $\rho:\Sigma_\red\to \CP^1$ that is a double covering, 
branched over 6 points.  $\Sigma_\red$
has a $\Z_2$ symmetry group, generated by a symmetry $\tau:\Sigma_\red\to\Sigma_\red$ that exchanges the two sheets of the covering.  
Concretely, $\Sigma_\red$ can be described by a hyperelliptic equation
\begin{equation}\label{zolg}y^2=P_6(u,v),\end{equation}
where $P_6$ is a homogeneous polynomial of degree 6 in homogeneous coordinates $u,v$ of $\CP^1$.  
In this description, $\rho$ is defined by forgetting $y$ and $\tau$ acts  by $y\to -y$.  
For $q$ a point in $\Sigma_\red$, we sometimes write $q'$ for $\tau(q)$ and say that $q$ and $q'$ are conjugate.
The condition that $\O(p_1+p_2)\cong K$ means that there is a holomorphic differential on $\Sigma_\red$ whose zeroes are $p_1$ and $p_2$.
A holomorphic differential on $\Sigma_\red$ has the form $\omega = (au+bv)(u\d v-v \d u)/y$, with constants $a,b$.  It vanishes
precisely when $au+bv=0$, a condition that defines a single point in $\CP^1$, but a conjugate pair of points in $\Sigma_\red$.
Since $p_1$ can be chosen arbitrarily and $p_2$ is then determined to be $\tau(p_1)$, the condition that $p_1$ and $p_2$ are conjugate
defines a divisor $\DD$ in the space of all pairs $p_1,p_2$.   For the case $g=2,r=1$, this is the bad set in the definition of the super period matrix.
As explained in detail in Appendix \ref{perturb}, the super period matrix has a pole with nilpotent residue
along $\DD$. 

This pole does not lead to additional contributions in the Ward identity (\ref{mono}) because  when the Ramond
divisors are associated to conjugate points $p_1,p_2$, the fermions of the matter system  have many zero-modes that cancel the potential
singularity due to the pole in the period matrix.  To understand this, we must recall some facts about the superstring spin field $S_A$ and 
its coupling to the matter
fermions of the RNS model. 

\subsection{Fermion Zero Modes}\label{fzero}
First we consider the case of uncompactified ten-dimensional spacetime.  The rotation group is $\SO(10)$, and its maximal torus is $U(1)^5$.
The weights of the spinor representation take the form $\veps_1,\veps_2,\dots,\veps_5$, where $\veps_i$ is the charge under the $i^{th}$ $U(1)$ and each
$\veps_i$ is $\pm 1/2$.  For a spinor of positive chirality, the number of weights $-1/2$ is even; we will assume that the spin field $S_A$ is a spinor of positive
chirality. 

The matter fermions of the heterotic string can be grouped as five complex fermions $\psi_i$ and their charge conjugates $\h\psi_i$, all of which are holomorphic fields.  
$\psi_i$ and $\h\psi_i$
have charge 1 and $-1$, respectively for the $i^{th}$ $U(1)$ in the maximal torus, and zero charge for the other $U(1)$'s.   The fields $\psi_i$ and
$\h\psi_i$ appear in a Dirac action $\int_{\Sigma_\red}\d^2z\, \h\psi_i\t\partial\psi_i$, and this tells us that if $\psi_i$ is a section of a line bundle $\L$,
then $\h\psi_i$ is a section of a conjugate line bundle $K\otimes \L^{-1}$.

 Suppose that these fermions
interact with spin fields placed at points $p_\alpha\in\Sigma_\red,~~\alpha=1,\dots,2r$, and let the weights of the spin field at $p_\alpha$ be
$\veps_{i,\alpha}=\pm 1/2$, $i=1,\dots ,5$.  In the most traditional formulation, one says that as $z\to p_\alpha$, $\psi_i(z)$ has a half-order zero or pole
at $z=p_\alpha$, depending on $\veps_{i,\alpha}$, with $\psi_i(z)\sim (z-p_\alpha)^{-\veps_{i,\alpha}}$.
In terms of complex geometry, this means roughly that $\psi_i$ is a section of $K^{1/2}\otimes_{\alpha=1}^{2r}\O(p_\alpha)^{\veps_{i,\alpha}}.$
But what  precisely is meant by the half-integral powers $\O(p_\alpha)^{\pm 1/2}$?  If (for some $i$) the $\veps_{i,\alpha}$ are all $+1/2$,
the meaning is simply that $\psi_i$ is a section of $\R^{-1}$ (which has an isomorphism $(\R^{-1})^2\cong K\otimes_\alpha \O(p_\alpha)$, so it is informally $K^{1/2}\otimes_\alpha
\O(p_\alpha)^{1/2}$).  In general, $\psi_i$ is a section of $\L_i=\R^{-1}\otimes_{\alpha|\veps_{i,\alpha}=-1/2}\O(-p_\alpha)$, and dually  $\h\psi_i$ is a section
of $K\otimes \L^{-1}\cong K\otimes \R\otimes_{\alpha|\veps_{i,\alpha}=-1/2}\O(p_\alpha)$.  These statements incorporate the usual assertions about zeroes
and poles of half-order, but in a way that is more natural in algebraic geometry.  By including in $\L_i$ a factor of $\O(-p_\alpha)$ whenever $\veps_{i,\alpha}=-1/2$, we ensure that in this case
(in the more informal language), $\psi_i$ has a half-order zero at $p_\alpha$ rather than a half-order pole.

In our application, 
there are just two Ramond insertions -- the operator $\S_A$ at one point $p$ and the operator $\V^B=\t\partial X^I \Gamma_I^{BC}\S_C$
at another point $q$. (To reduce clutter, in the following discussion we write $p$ and $q$ rather than $p_1$ and $p_2$ for the points with
Ramond insertions.)   We first discuss the case of superstring theory in $\RR^{10}$. 
It is convenient to pick  $\S_A$ to be the particular spin field whose  weights are all $1/2$, and to pick  $\V^B$ to have all weights $-1/2$.  This choice ensures
that the dilaton vertex operator
$\V_{\mathrm{dil}}$ does appear as a pole in the product $\S_A\,\V^B$.  Acting with a matrix $\Gamma_I$ (for any value of $I$) will
always flip precisely one weight.     So (looking at the definition (\ref{olor}) of $\V^B$) the component of $\V^B$ with all weights $-1/2$ is
a sum of terms, each proportional to a spin field $\S_C$ that has four weights $-1/2$ and one weight $+1/2$. As long as we are on $\RR^{10}$,
for counting fermion zero-modes, the different components
are equivalent.  We may as well look at the term
with  $\veps_{1,q}=1/2$ and $\veps_{i,q}=-1/2$, $i>1$.  

With those weights, $\psi_1$ is a section of $\R^{-1}$.
As we have seen above, the critical case that might give an unwanted contribution to the identity (\ref{mono}) is 
that $q=p'$ (that is, $q$ is conjugate to $p$), and moreover in this case, $\R^{-1}\cong K\cong \O(p+q)$.
So in the critical case,  $\psi_1$ is a section of $\R^{-1}\cong  K$ and dually  $\h\psi_1$ is a section of $K\otimes \R\cong \O$.  On the other hand, $\psi_i$ for $i>1$ is a section of
$\R^{-1}\otimes \O(-q)\cong \O(p)$.  Dually, $\h\psi_i$ for
$i>1$ is a section of $K\otimes \O(p)^{-1}\cong \O(q)$.

Now we can count fermion zero modes.  The line bundle $K$ has a two-dimensional space of holomorphic sections, while $\O$, $\O(p)$, and $\O(p')$
all have one-dimensional spaces of holomorphic sections.  So in the critical case $q=p'$, all $\psi_i$ and $\h\psi_j$ have 1 zero-mode, except that
$\psi_1$ has 2, making a total of 11 fermion zero-modes in all.

This is far too many zero modes for the locus $q=p'$ to be dangerous.  Before discussing what happens to the identity (\ref{mono}), let us simply
discuss how the correlation function $F_{\S_A\V^B}$, understood as a form on $\MM_{2,0,2}$ of codimension 1, behaves for $q\to p'$.
$\MM_{2,0,2}$ has odd dimension 3, so in conventional language, to compute $F_{\S_A \V^B}$, one inserts three picture-changing operators on the
worldsheet $\Sigma$.  Each picture-changing operator (PCO) contains a factor of the worldsheet supercurrent, which (for strings in $\RR^{10}$) is linear in the RNS fermions 
$\psi_i$ and $\h\psi_j$ and so can absorb one zero-mode.  So the PCO's remove 3 zero-modes, leaving 8.  Because of the 8 remaining zero-modes, $F_{\S_A\V^B}$ vanishes on the divisor $q=p'$.  To determine the order of its vanishing, we let $u$ be a local parameter on $\MM_{2,0,2}$ with a simple zero at $q=p'$.
At $u=0$, there are eight fermion zero-modes that are not lifted by the PCO's.  All of these modes are lifted away from zero for $u\not=0$ and generically
they are lifted to first order in $u$.  So if we write $\zeta_1,\dots,\zeta_8$ for the relevant modes that become zero-modes at $u=0$, the integral over those
modes near $u=0$ looks like
\begin{equation}\label{mondy}\int\d^8\zeta\,\exp(u m_{ij}\zeta_i\zeta_j)\sim u^4 \,{\mathrm{Pfaff}(m)},\end{equation}
where $m$ is an antisymmetric form that generically is nondegenerate and $\mathrm{Pfaff}(m)$ is its Pfaffian.   Thus $F_{\S_A\V^B}\sim u^4$ for $u\to 0$.  

The analysis of the identity (\ref{mono}) is more tricky than this, because in defining $F^*_{\S_A\V^B}$, we remove a factor $\pi^*(\mu)$  that  has a pole
 at $u=0$.  As we explain in section \ref{poleffect}, the effect of this is to cancel one power of $u$,
so that $F^*_{\S_A\V^B}\sim u^3$ for $u\to 0$.  In particular, $F^*_{\S_A\V^B}$ has no singularity that would contribute to the identity (\ref{mono}).

Now let us consider compactification to four dimensions.  The main difference is that the fields $\psi_j$, $\h\psi_j$, $j=3,4,5$ are not free fields on the worldsheet;
we cannot talk about their zero-modes or use them in a simple way to predict a zero of an amplitude. Also, we need to slightly modify the definition of the dilatino vertex
operator
 $\V^B$ that has all weights $-1/2$; 
 it is now the sum of two terms, each proportional to a spin field $S_C$ whose weights $\veps_i$ are all $-1/2$ except for either $\veps_1$ or $\veps_2$ (this ensures
that the part of $\V^B$ in the internal compact manifold is actually a holomorphic primary field of the appropriate dimension).
 In the counting of zero-modes, we should consider only
$\psi_i$ and $\h\psi_i$ for $i=1,2$, and the number of such modes, counted the same way as before, is  now 5 instead of 11.  Repeating the previous reasoning but with the smaller number of
zero-modes, we find that $F_{\S_A\V^B}$ is of order $u$ for $u\to 0$ and $F^*_{\S_A\V^B}$ is of order 1.    In particular, $F^*_{\S_A\V^B}$ has no singularity
at $u=0$  that could contribute in the identity (\ref{mono}).  Hence in general, with the d'Hoker-Phong procedure, the ``bulk'' contribution to the superstring measure vanishes in genus 2 after integrating over odd moduli and summing over spin structures, but before any bosonic integrations.

\subsection{Behavior of the Correlation Function}\label{poleffect}

Our goal here is to justify some of these statements by describing the behavior of $F^*_{\S_A\V^B}$ along the divisor $\frak D\subset \MM_{2,0,2}$ where the super period matrix has a pole.
It is helpful first to recall (following \cite{Verlindes}, and using the language explained in section 3.6 of \cite{Revisited}) the origin
of the picture-changing operators in worldsheet path integrals.  For each odd variable $\eta$, it is convenient to introduce also its differential $\d\eta$.
Then the insertion of a PCO at a point $p\in\Sigma_\red$ comes from an integral
\begin{equation}\label{zong}\int \D(\eta,\d\eta)\,\exp(\d\eta\beta(p)+\eta S_{z\theta}(p)), \end{equation}
where $\beta$ is the usual commuting antighost field of superstring perturbation theory and $S_{z\theta}$ is the worldsheet supercurrent.
The integral over $\d\eta$ gives a factor $\delta(\beta(p))$, and naively the integral over $\eta$ gives a factor of $S_{z\theta}$.  The product
is the PCO $\delta(\beta(p))S_{z\theta}(p)$.  However, although this treatment of the integral over $\d\eta$ is correct, there is more subtlety in the integral over $\eta$.
Because of the lack of a natural separation between even and odd variables on supermoduli space, there are in general, depending on precisely how
one parametrizes supermoduli space, additional $\eta$-dependent contributions hidden
in other parts of the worldsheet path integral, beyond those written in (\ref{zong}). We can expand $\exp(\eta S_{z\theta}(p))=1+\eta S_{z\theta}(p)$.
The $\eta S_{z\theta}(p)$ term leads after integrating over $\eta$
to an insertion at $p$ of the usual PCO $\delta(\beta(p))S_{z\theta}(p)$, but the ``1'' term leads to insertion at $p$ of the operator
$\delta(\beta(p))$, which we might call an incomplete PCO.  Locally, one can parametrize supermoduli space in such a way that the incomplete PCO's
can be ignored, but in general, globally either one cannot do this or doing this introduces other complications.  So it is best to keep track of the contributions
involving incomplete PCO's.  The only general constraint is that the worldsheet path
integral with insertion of an odd number of incomplete PCO's vanishes because of fermi statistics, so one can always assume the number of incomplete PCO's to be even.

Now we consider our problem of the influence of the poles of the super period matrix on $F^*_{\S_A\V^B}$.
Those poles will only affect the dependence of $F^*_{\S_A\V^B}$ on holomorphic variables, and in what follows we may as well  ignore
the antiholomorphic variables.   $\MM_{2,0,2}$ is a smooth supermanifold (or rather orbifold) of dimension $5|3$, say with even and odd coordinates $h_1,\dots,h_5|\eta_1,\dots,\eta_3$.
$F_{\S_A\V^B}$ is, from a holomorphic point of view, a smooth top form, so ignoring its dependence on antiholomorphic variables, it is
\begin{equation}\label{tombo}F_{\S_A\V^B}= f(h_i|\eta_j) [\d h_1\dots \d h_5|\d\eta_1\dots\d\eta_3]. \end{equation}
The function $f(h_i|\eta_j)$ can be expanded in powers of the $\eta_i$.  
When $F_{\S_A\V^B}$ is computed in superconformal field theory, there is a contribution, proportional to $\eta_1\eta_2\eta_3$,
 that can be evaluated using complete PCO's only.
As there are 3 odd moduli, this contribution involves 3 insertions of $S_{z\theta}$. 
 These insertions can absorb 3 of the 11 fermion zero-modes (for strings in $\RR^{10}$)
 described in section \ref{fzero}, leading to a contribution to $f$ that, as explained there, is of order $u^4$ near $u=0$.  $F_{\S_A\V^B}$
also has a contribution that is linear in the $\eta_i$, arising from a contribution evaluated
with 2 incomplete PCO's and only one insertion of  $S_{z\theta}$.  This insertion lifts only one fermion zero-mode, giving a contribution to $F_{\S_A\V^B}$ that
is of order $u^5$.  We can thus rewrite (\ref{tombo}) in the more detailed form
\begin{equation}\label{wombo}F_{\S_A\V^B}=\left(f_0(h_i)\eta_1\eta_2\eta_3+\sum_{j=1}^3 f_j(h_i)\eta_j\right) [\d h_1\dots \d h_5|\d\eta_1\dots\d\eta_3], \end{equation}
with
\begin{equation}\label{tomob}f_0\sim u^4,~~ f_j\sim u^5, ~j>0.\end{equation}
Naively, the term in (\ref{wombo}) that is linear in the  $\eta$'s is irrelevant both because it vanishes more rapidly for $u\to 0$
 and because, with some of the $\eta$'s missing,
one might think that this term would not contribute in the integration over odd moduli that goes into evaluating the boundary contributions in eqn. (\ref{mono}).

However, this reasoning is not valid, for reasons that are related in part to some of the usual subtleties of superstring perturbation theory.
To define $F^*_{\S_A\V^B}$, we are supposed to split off from $F_{\S_A\V^B}$ the
pullback of a volume form on $\M_2$.  Concretely, let $m_1,m_2,m_3$ be the three independent matrix elements of the pseudoclassical block of the super period
matrix.  Then we can define $F^*_{\S_A\V^B}$ by
\begin{equation}\label{mofto} F_{\S_A\V^B}=[\d m_1\d m_2\d m_3]\cdot  F^*_{\S_A\V^B}, \end{equation}
where $F^*_{\S_A\V^B}$ is a relative volume form along the fibers of $\pi:\MM_{2,0,2}\to \M_2$.  The reason that pulling out the factor of $[\d m_1\d m_2\d m_3]$
changes the order of vanishing at $u=0$ is that the $m_i$ have poles at $u=0$.  With only three odd variables $\eta_i$, since the $m_i$ are even, the residue
of such a pole can only be a bilinear expression in the $\eta_i$, and the products of the residues of the poles in different $m_i$ would vanish.  Accordingly,
we lose nothing essential if we assume that only $m_1$ has a pole, and that its residue is proportional to $\eta_1\eta_2$.  We can pick coordinates so that when
the $\eta_i$ vanish, the $m_i$ coincide with $h_i$, $i=1,\dots,3$.  Then the general form of the pole is
\begin{equation}\label{tobby}m_1=h_1+\frac{\eta_1\eta_2 g(h_1,\dots,h_5)}{u} \end{equation}
for some function $g$.  So we see that
\begin{equation}\label{obby} \d m_1=\d h_1\left(1-\frac{\eta_1\eta_2 g}{u^2}\frac{\partial u}{\partial h_1}\right)+\dots ,\end{equation}
where we write only  terms proportional to $\d h_1$ on the right hand side; we have also dropped a term $\partial_{h_1}g \eta_1\eta_2/u $, as it is less singular than the $\eta_1\eta_2/u^2$ term that we have kept.  
If in defining $F^*_{\S_A\V^B}$ in eqn. (\ref{mofto}), we were to split off
a factor of $[\d h_1\d h_2\d h_3]$, then $F^*_{\S_A\V^B}$ would have the same behavior for $u\to 0$ as $F_{\S_A\V^B}$.  However, instead we are supposed
to split off a factor of $[\d m_1\d m_2\d m_3]\sim [\d h_1\d h_2\d h_3](1-\eta_1\eta_2 g\partial_{h_1}u/u^2+\dots)$, and this gives
\begin{equation}\label{zobby}F^*_{\S_A\V^B}\sim \left(f_0(h_i)\eta_1\eta_2\eta_3+ f_3(h_i)\eta_3\left(1+\frac{\eta_1\eta_2 g\partial_{h_1}u}{u^2}\right)+\dots\right)[\d h_4\d h_5|
\eta_1\eta_2\eta_3],\end{equation} where we omit some terms that do not affect the argument.
Since $f_3\sim u^5$, it follows that $F^*_{\S_A\V^B}$ has a contribution proportional to $u^3\eta_1\eta_2\eta_3$.  Since this term is proportional to the product of all three $\eta$'s, it can
contribute to the surface integrals on the right hand side of eqn. (\ref{mono}).  The effect of the poles of the super period matrix together with the use of
incomplete PCO's has been to reduce by 1
the expected order of vanishing of $F^*_{\S_A\V^B}$ near $u=0$: it vanishes as $u^3$ rather than $u^4$.

With more odd variables, we can repeat this process.  For every pair of odd variables, the order of vanishing of $F^*_{\S_A\V^B}$ along $u=0$
is reduced by 1 (or the order of a pole is increased by 1) compared
to what one would expect from counting fermion zero-modes while ignoring the pole of the super period matrix.

\subsection{Extension To Genus Three}\label{genthree}

Much less is known about superstring scattering amplitudes for $g>2$.  
However, for $g=3$, one can still define a meromorphic projection $\pi:\MM_{3,+}\to\M_{3,\spin+}$ using the super period matrix.
This map behaves badly along hyperelliptic divisors in $\M_{3,\spin+}$ that are described in Appendix \ref{hyperelliptic}, and this
may ultimately prevent it from being very useful.  In the following, we simply avoid these issues by assuming that
$\Sigma_\red$ (which is kept fixed in the whole analysis) is not hyperelliptic.

The natural superstring measure on $\MM_{3,+}$ 
can be pushed forward to a measure on $\M_{3,\spin+}$ by integrating over the fibers of $\pi$.  Here we will show, generalizing the above arguments
for $g=2$, that in the case of superstring theory in $\RR^{10}$, the resulting measure on $\M_{3,\spin+}$ vanishes when summed over spin structures,
without any integration over bosonic moduli. 

For $g=3$, the condition on a pair of Ramond insertions such that $h^0(\R^{-1})>1$ was determined in section \ref{badset}.  In particular,
this condition defines a divisor $\DD$  in the space of all pairs   $p,q\in \Sigma_\red$. 

 It actually turns out that the counting of fermion zero-modes can be done
without using explicit knowledge of $\DD$.
  For strings in $\RR^{10}$, with
the same configuration of spin fields as in section \ref{specialcase}, $\psi_1$ is a section of $\R^{-1}$, $\h\psi_1$ is a section
of $K\otimes \R$, $\psi_i$ for $i>1$ is a section of $\R^{-1}\otimes \O(-q)$, and $\h\psi_i$ for $i>1$ is a section of $K\otimes \R\otimes \O(q)$.
We have to count the number of zero-modes of these fermion fields.

The Riemann-Roch theorem and Serre duality tell us that $h^0(\R^{-1})-h^0(\Sigma,K\otimes \R)=1-g+\mathrm{deg}\,\R^{-1}=1$.  Along the divisor $\frak D$, $h^0(R^{-1})$ is
at least\footnote{Here and in what follows, the minimum values are also the generic values.} 2, so  $h^0(K\otimes \R)$ is at least 1.   Hence along $\frak D$,
the  fields $\psi_1 $ and  $\t\psi_1$  have together at least  $2+1=3$ zero-modes.  For the fields $\psi_i,$ $i>1$, we have to replace $\R^{-1}$ by $\R^{-1}\otimes \O(-q)$.
This replacement imposes 1 condition on a holomorphic section of $\R^{-1}$ (it must vanish at $q$), so it reduces the number of zero-modes by at most 1
(and generically by 1).  So along $\frak D$, the number of $\psi_i$ zero-modes for $i>1$ is  always at least 1.  For $\R^{-1}$ of degree
$g=3$, so that $\R^{-1}\otimes \O(-q)$ has degree 2,  the  Riemann-Roch
theorem and Serre duality imply that $h^0(\R^{-1}\otimes \O(-q))-h^0(K\otimes \R\otimes \O(q))=1-g+2=0$,  or $h^0(\R^{-1}\otimes \O(-q))=
h^0(K\otimes \R\otimes \O(q))$, 
so that $h^0(K\otimes \R\otimes \O(q))$, which is
the number of $\h\psi_j$ zero modes for $j>1$,  is also
at least 1.

In sum, along $\frak D$, we always have for $g=3$ at least the same 11 fermion zero-modes that we found for $g=2$ 
(and generically the number is precisely 11).  We can then use the same
reasoning as in section \ref{specialcase} to analyze the behavior of $F_{\S_A\V^B}$ and $F^*_{\S_A\V^B}$ along $\frak D$.  The only difference
is that the number of odd moduli of $\MM_{3,0,2}$ is 5 instead of 3.  So to compute $F_{\S_A\V^B}$, we would use 5 PCO's instead of 3, with
the result that $F_{\S_A\V^B}$ has a zero of order 3 along $\frak D$, rather than the zero of order 4 that we found for genus 2.   
In going from $F_{\S_A\V^B}$ to $F^*_{\S_A\V^B}$, we can now lose two orders of vanishing (one for each pair of incomplete PCO's), so $F^*_{\S_A\V^B}$ has a simple zero along $\frak D$.

As before, supersymmetric compactification to four dimensions reduces the number of zero-modes from 11 to 5.  But now, with 5 PCO's, $F_{\S_A\V^B}$ need
not vanish along $\frak D$, and $F^*_{\S_A\V^B}$ may have a double pole.  So it appears that in a general supersymmetric compactification to four dimensions,
the bulk contribution to the genus 3 vacuum amplitude, defined using the projection $\MM_{3,+}\to \M_{3,\spin+}$ derived from the super period matrix,
does not necessarily vanish pointwise after summing over spin structures.

For any $g$, with precisely 2 Ramond punctures, Riemann-Roch and Serre duality can be used in the same way just described,
to show that if $h^0(\R^{-1})\geq 2$, then the RNS fermions associated to strings in $\RR^{10}$  have at least 11  zero-modes.
But for $g>3$, this  cannot be used in any obvious way to study the superstring measure.  The most basic  problem is that for $g>3$, the super period matrix generically does not obey the Schottky relations that are satisfied by the period matrix of an ordinary Riemann surface, and consequently
cannot be used to define a projection $\MM_{g,+}\to \M_{g,\spin+}$.  Also, as $g$ increases, the odd dimension of $\MM_{g,0,2}$  increases.
For  $g\geq 4$, even if we did have a holomorphic projection $\MM_{g,+}\to \M_{g,\spin+}$ with the properties that we have exploited,
the above reasoning  would allow a pole of $F^*_{\S_A\V^B}$ along $\frak D$, leading to no simple conclusion from the identity (\ref{mono}).

\appendix

\section{The Pole Of The Super Period Matrix For $r=1$}\label{perturb}

For the case of two Ramond punctures, that is $r=1$, we have found in section \ref{badset} that the exceptional set along which $h^0(\R^{-1})\geq 2$
is of codimension 1, and thus defines a divisor $\DD$ in the reduced space of $\MM_{g,0,2}$.     In section \ref{poleffect}, it was important to know
how the super period matrix $\h\Omega$ behaves along $\DD$.  We claim that the pseudoclassical block $\h\Omega_{g\times g}$ of the super period
matrix has a pole along $\DD$ with nilpotent residue.  To be more precise, the claim is that if $u$ is a local parameter on $\MM_{g,0,2,\red}$ with a simple
zero along $\DD$, and $\eta_1,\dots,\eta_s$ are the odd moduli, then $\h\Omega_{g\times g}$ has an expansion in which the leading term at $\eta=0$
is the classical period matrix $\Omega$ and the corrections are a series in $\eta_a\eta_b/u$:
\begin{equation}\label{holfo} \h\Omega=\Omega+\frac{w^{(2)}_{ab}\eta_a\eta_b}{u}+\frac{w^{(4)}_{abcd}\eta_a\eta_b\eta_c\eta_d}{u^2}+\dots . \end{equation}
Here $w_{ab}^{(2)},\, w_{abcd}^{(4)},\dots$ are functions on $\DD$, and less singular terms are omitted.   To deduce this formula, we simply use the D'Hoker-Phong expansion of the super period
matrix as a function of odd variables \cite{DPhagain}, slightly adapted to take into account the presence of Ramond punctures. 
   For each pair of
odd variables, the D'Hoker-Phong formula contains a fermion propagator (called $S(z,z')$ below) that behaves as $1/u$ because of the presence of zero-modes at $u=0$.
This accounts for the form of the expansion in eqn. (\ref{holfo}).

We will assume that the reader is familiar with the derivation of the D'Hoker-Phong expansion given in section 8.3 of \cite{Wittensurf}.  We will
essentially repeat this derivation, with some minor modifications to account for Ramond punctures.

Starting with a split super Riemann surface $\Sigma$, we want to give a smooth model for
its deformations associated to odd moduli.   In the absence of Ramond punctures, this is
done as follows (eqn. (8.17) of \cite{Wittensurf}).  Locally $\Sigma$ can be described by holomorphic
superconformal coordinates $z|\theta$ and a local antiholomorphic coordinate $\t z$; in the case that $\Sigma$
is split, we can take $\t z$ to be the complex conjugate of $z$.  A holomorphic function on $\Sigma$
is a function annihilated by $\partial_{\t z}$.  To deform the complex structure of $\Sigma$, we replace
$\partial_{\t z}$ by\footnote{In a more
complete treatment, we would include even deformations by adding to the right hand side an additional
term $h_{\t z}^z\partial_z+\frac{1}{2}\partial_z h_{\t z}^z\theta\partial_\theta$, where $h_{\t z}^z$ is a $(0,1)$-form
on $\Sigma_\red$ valued in $T$.  This is not necessary for extracting the singular behavior claimed in eqn. (\ref{holfo}).}
\begin{equation}\label{mortz}\partial'_{\t z}=\partial_{\t z}+\chi_{\t z}^\theta(\partial_\theta-\theta\partial_z), \end{equation}
where $\chi_{\t z}^\theta$ is a $(0,1)$-form on $\Sigma_\red$ valued in  $T^{1/2}$. ($T$ and $K$ will denote the holomorphic
tangent and cotangent bundles of $\SIgma_\red.$)  The expression $\chi_{\t z}(\partial_\theta
-\theta\partial_z)$ is an odd $(0,1)$-form on $\SIgma$ valued in odd superconformal vector fields.  

We need to find an analogous construction 
in the presence of Ramond punctures.  The first step is to reinterpret $\chi_{\t z}^\theta$ as a $(0,1)$-form on $\Sigma_\red$
that is valued in $\R$, which we characterize as in eqn. (\ref{elmonk}) as
 a line bundle on $\Sigma_\red$ with an isomorphism $\R^2\cong T(-p_1-\dots-p_{2r})$, 
 where $p_1,\dots,p_{2r}$ are the locations of Ramond punctures.  Equivalently, there is an isomorphism
$\R\cong \R^{-1}\otimes T(-p_1-\dots - p_{2r})$.   Yet another equivalent statement is that there is a homomorphism
\begin{equation}\label{minetrop}\ww:\R\to \R^{-1}\otimes T \end{equation}
which maps a section $s$ of $\R$ to a section $\ww(s)$ of $\R^{-1}\otimes T$ that vanishes at $p_1,\dots,p_{2r}$.

If $s$ is a section of $\R\to\Sigma_\red$, then $s\partial_\theta-\ww( s)\theta\partial_z$ is an odd superconformal vector
field on $\Sigma$.   (If the superconformal structure of $\Sigma$ is defined locally by $\varpi^*=\d z-z\theta\d\theta$,
then $s\partial_\theta-\ww(s)\theta\partial_z=s(\partial_\theta-\theta z\partial_z)$.)  With $s$ replaced by a $(0,1)$-form
$\chi_{\t z}^\theta$ valued in $\R$, we get the appropriate generalization of eqn. (\ref{mortz}) in the presence of Ramond
punctures:
\begin{equation}\label{yrfo}\partial_{\t z}'=\partial_{ \t z}+\chi_{\t z}^\theta\partial_\theta-\ww(\chi_{\t z}^\theta)\theta\partial_z.   \end{equation}
We expand $\chi_{\t z}^\theta$ in a basis $f_{a\,\t z}^\theta$ of $H^1(\Sigma_\red,\R)$, with coefficients the odd moduli $\eta_a$:
\begin{equation}\label{myrfo}\chi_{\t z}^\theta=\sum_{a=1}^{h^1(\R)}\eta_a \,f_{a\,\t z}^\theta. \end{equation}
In this way, the odd parameters $\eta_a,$ $a=1,\dots,h^1(\R)$ are used to deform the complex structure of $\Sigma$.

With this description of the perturbation that we are trying to make, it is actually straightforward to repeat the derivation in section 8.3 of \cite{Wittensurf}.
One basically just needs to replace  $\chi_{\t z}^\theta$  in some places by $\ww(\chi_{\t z}^\theta)$.

 A holomorphic 1-form $b(z)\d z$ on $\SIgma_\red$ corresponds to a section $\sigma=
b(z)\theta[\d z|\d\theta]$ of $\Ber(\Sigma)$.  If $\Sigma$ is split, the pseudoclassical block of its super period matrix coincides with
the classical period matrix of $\Sigma_\red$.  To compute the dependence of the super period matrix on odd moduli, we have to analyze
how a section $\sigma$ of $\Ber(\Sigma)$ changes when the odd moduli are turned on. 

 By analogy with eqn. (8.19) of \cite{Wittensurf}, a general section of $\Ber(\SIgma)$ in the presence of
the perturbation is 
\begin{equation}\label{morx}\h\sigma=\h\phi(\t z;z|\theta)\left[\d z+\ww(\chi_{\t z}^\theta)\theta\d\t z|\d\theta+\chi_{\t z}^\theta\d\t z\right]. \end{equation}
If $\h\sigma$ is understood as an integral form, then as in eqn. (8.22) of \cite{Wittensurf}, the condition for $\h\sigma$ to be holomorphic
is $0=\d\h\sigma$, where
\begin{equation}\label{donzo}\d\h\sigma=-\d\t z\d z\delta(\d\theta)\left(\partial_{\t z}\h\phi-\partial_z(\h\phi\,\ww(\chi_{\t z}^\theta)\theta)+
\partial_\theta(\h\phi\chi_{\t z}^\theta)       \right). \end{equation}
We expand 
\begin{equation}\label{durfo}\h\phi(\t z;z|\theta)= \hat\alpha(\t z;z)+\theta\h b(\t z;z). \end{equation}
For $\h\phi$ to be a section of $\Ber(\SIgma)$, $\hat\alpha$ should be a section of $K\otimes \R\cong \R^{-1}(-p_1-\dots -p_{2r})$, and 
$\h b$ should be a section of $K$.  
The condition $\d\h\sigma=0$ becomes a pair of equations, generalizing eqn. (8.23) of \cite{Wittensurf}:
\begin{align}\label{zebbo} \partial_{\t z}\h\alpha+\h b \chi_{\t z}^\theta & = 0 \cr
                                         \partial_{\t z}\h b-\partial_z\left(\h\alpha\, \ww(\chi_{\t z}^\theta)\right)& = 0. \end{align}
 Define an ordinary 1-form on $\Sigma_\red$:
 \begin{equation}\label{ebbo}\h\rho = \h b\,\d z+\h\alpha\,\ww(\chi_{\t z}^\theta) \,\d\t z. \end{equation}                                        
 The second equation in (\ref{zebbo}) says that   $\d\h\rho=0$.  The $A$- and $B$-periods of the closed holomorphic 1-form $\mu$ on $\Sigma$ that
 corresponds to    $\h\sigma$  are simply the ordinary $A$- and $B$-periods  of the ordinary 1-form $\h\rho$. (The proof of this
 statement can be found in section 8.3 of \cite{Wittensurf} and is unaffected by the existence of Ramond punctures.)  So to compute the super period
 matrix of $\Sigma$ as a function of odd moduli, we simply have to compute the periods of $\h\rho$.
 
 To do this, we have to solve the equations (\ref{zebbo}) as a function of the odd parameters $\eta_a$.  On considering the first of these
 equations, we immediately run into a problem.   Generically, this equation has no solution.  The obstruction lies in $H^1(\Sigma_\red,K\otimes \R)$,
 which generically (in the presence of $2r$ Ramond punctures) is of dimension $r$.  This obstruction reflects something that was explained
 in Appendix D.1 of \cite{Wittensurf}, and that was important in section \ref{counting} above.  In the presence of Ramond punctures, closed holomorphic
 1-forms on $\SIgma$ correspond to sections not of $\Ber(\Sigma)$, but of $\Ber'(\SIgma)$, where $\Ber'(\SIgma)$ is the sheaf whose sections
 are sections of $\Ber(\Sigma)$ that may have poles, with zero residue, along a Ramond divisor.  For our purposes, this means that in solving
 eqn. (\ref{zebbo}), we should allow $\h\alpha$ to have simple poles at $p_1,\dots,p_{2r}$.  Thus, it is a section not of $K\otimes \R$
 but of $K\otimes \R(p_1+\dots+p_{2r})\cong\R^{-1}$.   
 
 With $\h\alpha$ understood in this way, the first of eqns. (\ref{zebbo}) can be solved, but the solution is not unique: it is unique only
 modulo the possibility of adding to $\h\alpha$ an element of $H^0(\Sigma_\red,\R^{-1})$, which generically is of dimension $r$.  This non-uniqueness
 was to be expected; it reflects the fact that $\Ber'(\Sigma)$ has both odd and even holomorphic sections, with the odd sections
 having the form $\h\alpha [\d z|\d\theta]$ for $\h\alpha\in H^0(\Sigma_\red,\R^{-1})$.   When we deform a section $\sigma$ of $\Ber'(\SIgma)$ as a function of the odd moduli $\eta_a$, we are free to add to $\sigma$ a linear combination of the odd sections with $\eta_a$-dependent coefficients.

 The procedure in defining the super period matrix is to consider closed holomorphic 1-forms, or equivalently sections of $\Ber'(\Sigma)$, that
 have specified $A$-periods. 
  We start at $\eta_a=0$ with holomorphic 1-forms $\mu_j=b_j(z)\d z$ such that, on the ordinary Riemann
 surface $\Sigma_\red$, $\oint_{A^i_{\red}}\mu_j=\delta^i_j$.  The section $\sigma_j=\theta b_j(z)[\d z|\d \theta]$ of $\Ber'(\Sigma)$ then automatically has
 vanishing fermionic $A$- and $B$-periods.  We  deform $\sigma_j$ as a function of the odd moduli $\eta_a$ to get a section $\h\sigma_j$
 of $\Ber'(\Sigma)$ such that $\oint_{A^i}\h\sigma_j=\delta^i_j$, and $\h\sigma_j$ has vanishing fermionic $A$-periods.  As in the classical
 theory, it is not possible to  constrain
 the bosonic or fermionic $B$-periods of $\hat\sigma_j$; these make up the $g\times g$ and $g\times r$ blocks of the super period matrix.

In particular,
the vanishing of the fermionic $A$-periods of $\h\sigma_j$  will make the solution for
 $\h\alpha$ unique.
 This condition means that $\h\alpha$ is not an arbitrary section of $\R^{-1}$ but (in the notation
 of eqn. (\ref{melz})) obeys \begin{equation}\label{refo}\h\alpha(w_{2\zeta-1})+\sqrt{-1}\,\h\alpha(w_{2\zeta})=0,
  ~~~\zeta=1,\dots,r. \end{equation}  
 Away from a bad locus in the reduced space of $\MM_{g,0,2r}$ on which the super period matrix develops a singularity, there is no holomorphic
 section of $\R^{-1}$ that satisfies (\ref{refo}),  but there is such a section if we allow a pole at some point $z'\in\Sigma_\red\backslash \{p_1,\dots,p_{2r}\}$.
 This means that there is a unique solution $S(z,z')$ of the equation
 \begin{equation}\label{morfo}\partial_{\t z}S(z,z')=2\pi \delta^2(z,z'). \end{equation}
 (The delta function is defined by $\int \d^2 z \,\delta^2(z,z')=1$, where $\d^2z =-i\d \t z \,\d z$.)
  $S$ plays the role in the presence of Ramond
 punctures that the ordinary Dirac propagator plays in their absence.  We will describe it more precisely momentarily.
We can express $\h\alpha$ in terms of $\h b$:
 \begin{equation}\label{ubeklin} \h\alpha(\t z;\neg z)=-\frac{1}{2\pi}\int_{\Sigma'_\red} S(z,z')  \chi_{\t z'}^\theta(\t z';\neg z'),\h b(\t z';\neg z')\d^2 z'.\end{equation}
  And hence we can write an equation for $\h b$ only:
\begin{equation}\label{ueklin}\partial_{\t z}\h b(\t z;\neg  z)=-\frac{1}{2\pi}\partial_z \int_{\Sigma_\red'} \xi(\chi_{\t z}^\theta(\t z;\neg z) ) S(z,z') 
\chi_{\t z'}^\theta(\t z';\neg z')\,\h b(\t z';\neg z') \d^2 z' \end{equation}

However, we should explain better what sort of geometric object is $S(z,z')$.  In its dependence on $z$, it is a section of $\R^{-1}$ that
is constrained by eqn. (\ref{refo}), but what about its dependence on $z'$?   Since the answer is a little tricky and involves a slightly
exotic use of duality, we first give an example.
We take $\Sigma$ to be the complex $z$-plane with Ramond divisors at $z=0$ and $z=\infty$, as in eqn. (\ref{zeldo}).  If as in that discussion
we write $\d\theta$ for a section of $\R^{-1}$ that has fermionic periods 1 and $\sqrt{-1}$ at $z=0$ and $z=\infty$, then
\begin{equation}\label{monkey}S(z,z')=\d\theta\boxtimes \d\theta' \,\frac{z+z'}{2(z-z')}.  \end{equation}
(The symbol $\boxtimes$ just denotes a tensor product of forms on two different factors of $\Sigma$, one parametrized by $z|\theta$
and one by $z'|\theta'$.)
The point of this formula is that the function $(z+z')/2(z-z')$ equals $-\frac{1}{2}$ or $+\frac{1}{2}$ at $z=0$ or $z=\infty$, and hence $S(z,z')$
has fermionic periods $-\frac{1}{2}$ and $+\frac{1}{2}\sqrt{-1}$ at $z=0$ and $z=\infty$.  The relative minus sign means that  while $\d\theta$ has a vanishing fermionic $B$-period,
$S(z,z')$ has a vanishing fermionic $A$-period in its dependence on $z$, as desired.  

We note from eqn. (\ref{monkey}) that $S(z,z')$ is odd under $z\leftrightarrow z'$, and in particular that it is the same sort of geometric
object in each variable.  This is what we want to explain in general.  We will use a physical language.
 If we are given a pair of fermi fields $\psi, \,\t\psi$ that are sections of
Serre dual line bundles $\L$ and $K\otimes \L^{-1}$, where $\L$ has degree $g-1$ and $h^0(\L)=0$, 
then we would have a Dirac action $I=\frac{1}{2\pi}\int \psi\partial_{\t z}
\t\psi$, with a fermion propagator $S(z,z')$ obeying eqn. (\ref{morfo}).  It would be a section of $\L\boxtimes K\otimes\L^{-1}\to \SIgma\times \Sigma$
(with a simple pole on the diagonal of residue 1).    In our case, $S(z,z')$ as a function of $z$ is a section of the line bundle $\R^{-1}$ of
degree $g-1+r$, but it obeys the $r$ constraints (\ref{refo}).  To write an action describing this situation, we take $\psi$ and $\t\psi$
to be sections of $\R^{-1}$ and $K\otimes \R$, respectively, but with  Lagrange multipliers
that enforce the constraints:
\begin{equation}\label{tolf} I=\frac{1}{2\pi}\int \psi\partial_{\t z}
\t\psi+\sum_{\zeta=1}^r c_\zeta \left(\psi(p_{2\zeta-1})+\sqrt{-1}\psi(p_{2\zeta})\right). \end{equation}
Varying with respect to $\psi$, the equation of motion for $\t\psi$ is
\begin{equation}\label{merof} \frac{1}{2\pi}\partial_{\t z}\t\psi(z)=\sum_{\zeta=1}^r c_\zeta\left(\delta^2(z,z_{2\zeta-1})+\sqrt{-1}\delta^2(z,z_{2\zeta}\right).\end{equation}
Thus, $\t\psi$ has poles at $p_1,\dots,p_{2r}$, and we can view it as a section of $K\otimes \R(p_1+\dots+ p_{2r})\cong \R^{-1}$.
But the $2r$ poles are not independent; their residues are determined by the $r$ Lagrange multipliers $c_\zeta$.  The resulting relations
between the residues mean precisely that as a section of $\R^{-1}$, $\t\psi$ obeys the constraints (\ref{refo}) and thus that it is the same
sort of geometric object as $\psi$.  Thus we can view $S(z,z')$ in eqn. (\ref{morfo}) as a section of $\R^{-1}\boxtimes\R^{-1}$ that obeys
the conditions (\ref{refo}) in each variable and has a simple pole on the diagonal.  $S(z',z)$ obeys all of the same conditions, but with an
opposite residue for the pole on the diagonal, so
\begin{equation}\label{tofu}S(z',z)=-S(z,z'). \end{equation}
This is important in the symmetry of the formula (\ref{golly}) below for the first correction to the super period matrix.

However, if we want to view $S(z,z')$ as a section of $\R^{-1}\boxtimes \R^{-1}$ rather than of $\R^{-1}\boxtimes K\otimes \R(p_1+\dots+p_{2r})$,
we have to replace $\chi_{\t z'}^\theta$ in eqns. (\ref{ubeklin}) and (\ref{ueklin}) with $\xi(\chi_{\t z'}^\theta)$, to give
 \begin{equation}\label{beklin} \h\alpha(\t z;\neg z)=-\frac{1}{2\pi}\int_{\Sigma'_\red} S(z,z') \xi( \chi_{\t z'}^\theta(\t z';\neg z'))\,\h b(\t z';\neg z')\d ^2z'\end{equation}
and
\begin{equation}\label{eklin}\partial_{\t z}\h b(\t z;\neg  z)=-\frac{1}{2\pi}\partial_z \int_{\Sigma_\red'} \xi(\chi_{\t z}^\theta(\t z;\neg z) ) S(z,z') 
\xi(\chi_{\t z'}^\theta(\t z';\neg z'))\,\h b(\t z';\neg z') \d^2 z'. \end{equation}

Starting with a choice of $\h b$ at $\eta_a=0$,   eqn. (\ref{eklin}) can be solved  iteratively to determine $\h b$ as a function of the $\eta_a$.
To make the solution unique, we need a condition on the $A$-periods of the ordinary 1-form $\h\rho$.  To find the right condition,
recall that to compute the super period matrix,
we  start with holomorphic 1-forms $\mu_j=b_j(z)\d z$ on $\Sigma_\red$ with canonical $A$-periods, $\oint_{A^i_\red}b_j(z)\d z=\delta^i_j$.
We promote them to $\eta_a$-dependent holomorphic sections $\h\sigma_j$ on $\Sigma$, such that, with $\h\rho_j$ defined by
eqns. (\ref{morx}), (\ref{durfo}) and (\ref{ebbo}),
\begin{equation}\label{worry}\oint_{A^i_\red}\h\rho_j=\delta^i_j. \end{equation}
This condition on bosonic $A$-periods  makes the solution of eqn. (\ref{eklin})  unique.

The final computation of the super period matrix $\h\Omega$ proceeds rather as  in section 8.3 of \cite{Wittensurf}.
The $g\times g$ block of $\h\Omega$ is defined as
\begin{equation}\label{wonx}\hat\Omega_{ij}=\oint_{B_j}\hat\sigma_i=\oint_{B_{j,\red}}\hat\rho_i.\end{equation}
The difference between  $\hat\Omega_{ij}$ and the classical period matrix $\Omega_{ij}$ is
\begin{equation}\label{onx}\hat\Omega_{ij}-\Omega_{ij}=\oint_{B_{j,\red}}\left(\hat\rho_i-\mu_i\right)=\oint_{B_{j,\red}}\h\rho_i',\end{equation}
where $\h\rho_i'=\h\rho_i-\mu_i$.  
Riemann's bilinear relations say that if $\kalpha,\kbeta$ are closed 1-forms on the ordinary Riemann surface $\Sigma_\red$, then
\begin{equation}\label{itzo}\int_{\Sigma_\red}\kalpha\wedge \kbeta=\sum_i\left(\oint_{A^i_\red}\kalpha\oint_{B_{i,\red}}\kbeta-\oint_{B_{i,\red}}\kbeta\oint_{A^i_\red}\kalpha\right).\end{equation}
Taking  $\kalpha=\mu_j=b_j(z)\d z$, $\kbeta=\hat\rho'_i$, and remembering that $\mu_j$ is of type $(1,0)$ on the ordinary Riemann surface $\Sigma_\red$, so that
the $(1,0)$ part of  $\hat\rho'_i$ does not contribute to $\mu_j\wedge\hat\rho'_i$,
we learn that 
\begin{equation}\label{monix}\hat\Omega_{ij}-\Omega_{ij}=\int_{\Sigma_\red} \mu_j\wedge\hat\rho'_i
=\int_{\SIgma_\red}\mu_j\wedge \hat\alpha_i\ww( \chi_{\t z}^\theta)\,\d\t z.\end{equation}
In turn, we can use (\ref{beklin}) to eliminate $\hat\alpha_i$:
\begin{equation}\label{delfom}\hat\Omega_{ij}-\Omega_{ij}=-\frac{1}{2\pi}\int_{\Sigma_\red\times \SIgma'_\red}\mu_j(z)\xi(\chi_{\t z}^\theta(\t z;z))\d\t z\,  S(z,z')
 \xi(\chi_{\t z}^\theta(\t z';\neg z'))\,\h b_i(\t z';\neg z')\d ^2z'.\end{equation}
When $\hat\Omega_{ij}-\Omega_{ij}$ is expanded in powers of the $\eta_a$'s, the lowest order term, which we call $\hat\Omega^{(2)}_{ij}$,
 is quadratic.  It
can found by just replacing $\hat b_i(\t z';\neg z')$ in the last formula by $b_i$ and writing $b_i\d^2z =-i \d\t z \,\mu_i$:
\begin{equation}\label{golly} \hat\Omega^{(2)}_{ij} =\frac{i}{2\pi}\int_{\Sigma_\red\times\Sigma'{}_\red}\mu_j(z)\xi(\chi_{\t z}^\theta(\t z;z))\d\t z\,S(z,z')\xi(\chi_{\t z'}^\theta(z'))\d\t z'\,\mu_i(z').\end{equation}
This is the analog of the D'Hoker-Phong formula in the absence of Ramond punctures.  
Higher order terms can be evaluated by using (\ref{eklin}) to express $\hat b_i$ as
a polynomial in the $\eta$'s.

Now we can explain the form of the expansion in eqn. (\ref{holfo}).   The kernel $S(z,z')$ develops a pole as a function of the moduli of $\Sigma$
when the $\partial_{\t z}$ operator, acting on sections of $\R^{-1}$ that satisfy eqn. (\ref{refo}), has a non-trivial kernel.  This is the bad set
discussed in section \ref{badset}. If $u$ is a local parameter with a simple zero on the bad set, then $S(z,z')\sim 1/u$ near $u=0$.
When $\h\Omega_{ij}-\Omega_{ij}$ is evaluated by solving eqn. (\ref{eklin}) iteratively for $b$ and substituting in eqn. (\ref{delfom}), each
pair of odd moduli is accompanied by a factor of $S(z,z')$.  This leads to the behavior claimed in eqn. (\ref{holfo}).

The meaning of the poles at $u=0$ is, however, quite different for $r=1$ or $r>1$.  For $r=1$, the choice of fermionic $A$-periods is
essentially unique,  as explained in section \ref{tworp}; the condition $h^0(\R^{-1})>r$ defines a divisor $\DD$; and the poles of $\h\Omega_{ij}$ along
this divisor do not depend on any arbitrary choices.  By contrast, for $r>1$, the definition of $\h\Omega_{ij}$ does depend on a somewhat
arbitrary choice of fermionic $A$-periods, and $\h\Omega_{ij}$ develops poles along a divisor on which this choice breaks down.  These
are the poles described in the above formula.  The
condition $h^0(\R^{-1})>r$ for $r>1$ defines a set of complex codimension greater than 1, and is not the locus of poles in $\h\Omega_{ij}$.

\section{A Note On The Behavior At Infinity In Genus Two}\label{infnote}

As explained in the introduction, in general the procedure of D'Hoker and Phong to compute the two-loop vacuum amplitude
must be supplemented by a correction at infinity.

 The explanation  given in \cite{Wittenmore} for the need for this correction was the following.  The correction is associated to the splitting of
  a genus 2 super Riemann surface $\Sigma$
 with even spin structure  to two genus 1 components each with even spin structure.  So consider a partial compactification\footnote{We introduce this
 partial compactification because $\pi$ does have a pole at infinity if one tries to extend it over the full Deligne-Mumford compactification of $\MM_{2,+}$.  The pole arises
 on the degeneration in which $\Sigma$ decomposes to two genus 1 components each with odd spin structure.  Because of fermion zero-modes, this pole in $\pi$ does not
 lead to any correction to the D'Hoker-Phong procedure for analyzing the vacuum amplitude.}
 in which one allows this type of degeneration, thus adding a divisor $\frak T$ to $\MM_{2,+}$.
Although the holomorphic splitting $\pi:\MM_{2,+}\to\M_{2,\spin+}$ used by D'Hoker and Phong
does extend to a holomorphic map between the corresponding partially compactified spaces,  the extended map does not restrict to a splitting of the divisor $\frak T$.
This leads to a slight mismatch between what one gets by first integrating over the fibers of $\pi$ 
 and the general formalism of superstring perturbation theory.   
 
 We will here give an alternative explanation in the language of picture-changing operators rather than super Riemann surface
 theory.  In Appendix \ref{genone},  we recall some elementary facts about Riemann surfaces of genus 1. Then in Appendix \ref{dph}, we recall  
 what the D'Hoker-Phong procedure means when expressed in terms of picture-changing operators (PCO's) and analyze how this procedure behaves near the relevant separating degeneration.
 
 \subsection{Spin Structures In Genus 1}\label{genone}

Let $\Sigma_1$ be a Riemann surface of genus 1.  Picking a point $p\in\SIgma_1$ as the ``origin,'' $\Sigma_1$ becomes
an elliptic curve.  Although the canonical bundle $K$ of $\Sigma_1$ is trivial, a square root $K^{1/2}$ of $K$ may not be trivial.
An even spin structure on $\Sigma_1$ corresponds to the case that $K^{1/2}$ is a non-trivial line bundle of order 2, namely
$K^{1/2}=\O(-p+q)$, where $q$ is one of the three non-zero points of order 2 on $\Sigma_1$.  

The Dirac propagator $S(y,z)$ is a section of $K^{1/2}\boxtimes K^{1/2}\to \Sigma_1\times \SIgma_1$ with a simple pole of residue
1 on the diagonal.  We will need to understand the propagator for the case that $y=p$.  $S(p,z)$ is, as a function of $z$, a nonvanishing section
of $K^{1/2}=\O(-p+q)$ that has a simple pole at $p$ (and no other singularities).  Differently put, it is a nonzero section of $\O(q)$.  But
a nonzero section of $\O(q)$ vanishes at $q$ and nowhere else.  Thus, $S(p,z)$ vanishes precisely at $z=q$.

Suppose that $\Sigma_1$ is the reduced space of a super Riemann surface $\SIgma$ with one NS puncture.  
In the picture-changing formalism,
we represent an NS puncture on $\SIgma$ as a point in $\SIgma_1$, which because of the translation symmetries of $\Sigma_1$ we may as well take to be $p$, and we represent the
odd modulus of $\SIgma$ by including a PCO at some point $u\in \SIgma$.  The only constraint on $u$
is that to avoid a ``spurious singularity,'' we want $H^0(\SIgma_1,T^{1/2}(-p)\otimes \O(u))=0$. (Here $T^{1/2}(-p)$ is the sheaf of odd
superconformal vector fields that vanish at $p$, and the condition to avoid a spurious singularity is that this sheaf has no section
whose only singularity is a simple pole at the position $u$ of the PCO.)  But $T^{1/2}(-p)\cong \O(-q)$ so we want $H^0(\Sigma_1,\O(-q+u))=0$.  This is true if and only if $u\not=q$.  In other words, we may place the PCO anywhere except at $q$ without running into a spurious singularity.

\subsection{The D'Hoker-Phong Procedure}\label{dph}

Now let $\Sigma_2$ be a Riemann surface of genus 2 with even spin structure.  We view it as the reduced space of a genus 2 super Riemann
surface $\SIgma$.   A family of $\SIgma$'s depending on a full set of odd parameters -- namey $2g-2=2$ of them -- can be constructed
by inserting PCO's at two points $u,v\in\Sigma_2$.  In this language, the D'Hoker-Phong procedure amounts to the following:  one should
choose the points $u,v$ so that the Dirac propagator connecting them vanishes:  $S(u,v)=0$. (To obey $S(u,v)=0$, we may pick any $u$
and for given $u$ there are then two choices of $v$.)  The D'Hoker-Phong formula for the dependence
of the super period matrix on odd moduli (see \cite{DPhagain} and also \cite{Wittensurf}, section 8.3, or eqn. (\ref{golly}) above) 
shows that we get a family of
super Riemann surfaces whose super period matrix does not depend on the odd moduli if we include odd moduli by placing PCO's at
any points $u,v$ satisying $S(u,v)=0$, and avoiding spurious singularities.   (Generic points satisfying $S(u,v)=0$ do avoid spurious
singularities.)
This means in particular that as long as we require $S(u,v)=0$ and avoid spurious singularities, the PCO formalism will give
an answer that does not depend on the specific choice of $u$ and $v$.  This is the D'Hoker-Phong procedure expressed in terms of PCO's. 
Following this procedure will certainly give a unique, globally-defined answer.  There cannot be any global obstruction to following
this procedure, because $u$ and $v$ do not have to vary continuously; we can use different pairs $u,v$ in different regions of moduli
space.  

\begin{figure}
 \begin{center}
   \includegraphics[width=3in]{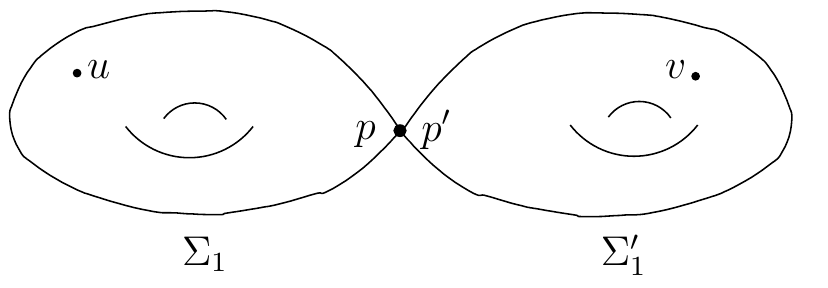}
 \end{center}
\caption{\small A singular genus 2 surface is made by gluing $p\in \SIgma_1$ to $p'\in\SIgma_1'$.  PCO's are inserted at $u\in\Sigma_1$
and $v\in\Sigma_1'$.} \label{twosplit}
\end{figure}

Now consider what happens when $\Sigma_2$ degenerates to a pair of genus 1 components $\Sigma_1$ and $\Sigma_1'$, each
with even spin structure.  $\Sigma_2$ is constructed by gluing a point $p\in\Sigma_1$ to a point $p'\in \Sigma_1'$ (fig. \ref{twosplit}).
The spin bundle of $\Sigma_1$ is $\O(-p+q)$ for some $q\in\SIgma_1$, and the spin bundle of $\Sigma_1'$ is $\O(-p'+q')$ for
some $q'\in \Sigma_1'$.  

The general formalism of superstring perturbation theory, when expressed in terms of PCO's, tells us that in the limit that $\Sigma_2$
degenerates as described in the last paragraph, we must ensure that 
the PCO's are in opposite branches, say $u\in\Sigma_1$, $v\in\Sigma_1'$.  (For example, see section 6.3.6 of \cite{Revisited}.)
 Moreover, to avoid a spurious singularity, we need $u\not=q$ and $v\not=q'$.

We will now see that there is a conflict between the D'Hoker-Phong procedure and the conditions stated in the last paragraph.  That is why
the D'Hoker-Phong procedure requires a correction at infinity.

If $\Sigma_2$ is a singular surface made by gluing together $\Sigma_1$ and $\Sigma_1'$, then the  Dirac propagator $S(u,v)$ is identically 0 for $u\in \Sigma_1$, $v\in\Sigma_1'$.  However, we really want to know what happens if
we impose the D'Hoker-Phong condition $S(u,v)=0$  away from the degeneration limit, and then let $\Sigma_2$ degenerate.  
If we deform a singular surface $\Sigma_2$ made by gluing at $p$ and $p'$  into a smooth surface with a small deformation parameter $\varepsilon$, then to lowest order in $\varepsilon$,
the propagator $S(u,v)$ with $u$ and $v$ on opposite branches is
\begin{equation}\label{dolfo}S(u,v)\sim \varepsilon S(u,p)S(p',v)+\O(\varepsilon^2).  \end{equation}
(For example, see eqn. (3.49) of \cite{Wittenmore}.)

As long as $\Sigma_2$ is smooth, there is no difficulty in ensuring that $S(u,v)=0$ and avoiding spurious singularities.  Now let us consider
when happens as $\varepsilon\to 0$.  Trying to avoid a spurious singularitiy, we take $u$ to not approach the point $q\in \Sigma_1$
as $\varepsilon\to 0$.   But then as we know from Appendix \ref{genone}, $S(u,p)\not\to 0$ for $\varepsilon\to 0$.  Eqn.  (\ref{dolfo}) then implies that to make $S(u,v)$
identically 0 for all $\varepsilon$, we will have to have $v\to q'$ for $\varepsilon\to 0$ in order to make $S(p',v)$ vanish. (Otherwise, $S(u,v)$ is of order $\varepsilon$
for small $\varepsilon$ and cannot vanish identically.)  But this means that for $\varepsilon\to 0$, we will
meet a spurious singularity after all.

In short, if we follow the D'Hoker-Phong procedure, then the divisor in which $\Sigma$ splits to a pair of genus 1 components with
even spin structure is a locus of spurious singularities.  That is why in general this procedure needs to be supplemented by adding
a correction term  supported on this divisor.

In this analysis, since we were not able to avoid a spurious singularity at infinity, we did not gain much by placing $u$ and $v$
on opposite branches of $\Sigma_2$ in the degeneration limit.  We could just as well have chosen $u$ and $v$ to be both in, say, $\Sigma_1$,
in the limit that $\Sigma_2$ degenerates.  For this, we could take $u$ to be a generic point in $\Sigma_1$ and then pick $v\in\Sigma_1$ such
that $S(u,v)=0$.  The correction at infinity can be computed by comparing this procedure to the general formalism of superstring perturbation theory.
 This can possibly be the basis for justifying the procedure of \cite{Sentwo}.

\section{The Hyperelliptic Locus In Genus Three}\label{hyperelliptic}

\subsection{The Purely Bosonic Case}\label{bosonic}

A hyperelliptic Riemann surface $\Sigma_3$ of  genus 3 can be described in affine coordinates by an equation
\begin{equation}\label{hypo}y^2=\prod_{a=1}^8(x-e_a). \end{equation}
Thus $\Sigma_3$ is a double cover of $\CP^1$ (parametrized by $x$ with a point at infinity added) with 8 branch points $e_1,\dots,e_8$.
The hyperelliptic involution acts by $\tau:y\to -y$, and acts as $-1$ on the three linearly independent holomorphic differentials
\begin{equation} \omega_t=\frac{\d x}{y}x^t,~~~t=0,1,2. \end{equation}
Dually, $\tau$ acts as $-1$ on $H_1(\SIgma_3,\Z)$ and hence on all $A$- and $B$-periods.  Accordingly the $3\times 3 $ period matrix
is even under $\tau$.

Modulo the action of $SL(2,\C)$ on $\CP^1$, this family of genus 3 hyperelliptic curves depends on 5 complex parameters (the action of
$SL(2,\C)$ can be used to fix 3 of the 8 branch points $e_a$).  These 5 moduli
are clearly $\tau$-invariant.
A genus 3 Riemann surface has  altogether $3g-3=6$ moduli, so there is 1 additional modulus $\veps$ that is odd under $\tau$.     
We can confirm this by examining
the quadratic differentials, which are dual to the infinitesimal deformations of $\Sigma_3$.  There are 5 even ones
\begin{equation}\label{moddo}\left(\frac{\d x}{y}\right)^2 x^t, ~~ 0\leq t\leq 4,\end{equation}
and 1 odd one
\begin{equation}\label{oddo}\frac{(\d x)^2}{y}. \end{equation}

A genus 3 Riemann surface is parametrized locally by its $3\times 3$ symmetric period matrix $\Omega_{ij}$.
Since hyperelliptic curves of genus 3 are a five-parameter family, they span a codimension 1 subspace of period matrices.  
Let $q$ be a holomorphic function of the matrix elements $\Omega_{ij}$ with a simple zero along the subspace that parametrizes
hyperelliptic Riemann surfaces.  Since all matrix elements of $\Omega_{ij}$ are $\tau$-invariant, $q$ in particular 
is an even function of $\veps$, and we can normalize $q$ and/or $\veps$ so that the relation
between them is simply
\begin{equation}\label{dobb}q=\veps^2. \end{equation}

\subsection{Spin Structures}\label{spinstr}

Now suppose that $\Sigma_3$ is the reduced space $\Sigma_\red$ of a split super Riemann surface $\Sigma$, and in particular is endowed
with a spin structure.  A genus 3 surface has $\frac{1}{2}(2^{2g}-2^g)=28$ odd spin structures and $\frac{1}{2}(2^{2g}+2^g)=36$ even ones.
A spin structure is a line bundle $\L$ with an isomorphism $\psi:\L\otimes \L\cong K$.  Such a line bundle can be characterized by specifying
$\psi(s\otimes s)$ for some meromorphic section $s$ of $\L$. Here  $\psi(s\otimes s)$ will have precisely the same zeroes
and poles as $s$, but with twice the multiplicity, so in particular it has zeroes and poles of even multiplicity only.  Often we write $K^{1/2}$ for $\L$.

 For example, to define an odd spin structure, we pick a pair of branch points $e_a$,
$e_b$,  and characterize $\L$ by saying that it has a section $s$ with 
\begin{equation}\label{murr} \psi(s\otimes s)=\frac{\d x}{y}(x-e_a)(x-e_b). \end{equation}
The differential $(\d x/y)(x-e_a)(x-e_b)$ has double zeroes at $x=e_a,\,y=0$ and at $x=e_b,\,y=0$, and no other zeroes or poles.
So $s$ has simple zeroes at those two points and no poles.  In particular, $s$ is holomorphic.   We can describe this more intuitively by writing
\begin{equation} \label{urr} s=\sqrt{\frac{\d x}{y}(x-e_a)(x-e_b)}.\end{equation}
Up to a constant multiple, $s$ is the only holomorphic
section of $\L$.  So $h^0(\L)=1$ and $\L$ is an odd spin structure.  The 28 odd spin structures on $\Sigma_\red$ are all obtained by this construction,
with the $8\cdot 7/2=28$ possible choices of the pair $e_a,\, e_b$.  Obviously, these 28 choices are permuted transitively by permutations of the  branch
 points $e_a$.
So in the moduli space $\M_{3,\spin-}$ that parametrizes a genus 3 surface with an odd spin structure, the hyperelliptic surfaces form an irreducible
divisor $\DD_-$.

By contrast, there are two essentially different types of even spin structure on $\Sigma_\red$, corresponding to whether $h^0(\L)$ equals 0 or 2.
There is only 1 spin structure with $h^0(\L)=2$.  The space $H^0(\SIgma_3,\L)$ is spanned by sections $s$, $s'$ that obey
$\psi(s\otimes s)=\d x/y$, $s'=xs$ ($s$ has simple zeroes at two points in $\Sigma_\red$ lying above $x=\infty$ and $s'$ has simple
zeroes at two points lying above $x=0$).    More informally, we write
\begin{equation}\label{delfo}s=\sqrt{\frac{\d x}{y}}. \end{equation}   Since there is just one spin structure with $h^0(\L)=2$, genus 3 surfaces
with such a spin structure are parametrized by an irreducible divisor $\DD\subset \M_{3,\spin+}$. 

An example of an even spin structure with $h^0(\L)=0$ is given by assuming that $\L$ has a meromorphic section $s$
with 
\begin{equation}\label{dundo} \psi(s\otimes s)=\frac{\d x}{y}\frac{(x-e_1)(x-e_2)(x-e_3)}{(x-e_4)}. \end{equation}
Informally,
\begin{equation}\label{Pundo}s=\sqrt{\frac{\d x}{y}\frac{(x-e_1)(x-e_2)(x-e_3)}{(x-e_4)}}.\end{equation}
Such an $\L$ has no holomorphic section.  How many choices are there of such $\L$'s?  We get an isomorphic line bundle if
we exchange $e_4$ with, say, $e_3$, since this can be accomplished by replacing $s$ by $s'=s(x-e_4)/(x-e_3)$:
\begin{equation}\label{dundox} \psi(s'\otimes s')=\frac{\d x}{y}\frac{(x-e_1)(x-e_2)(x-e_4)}{x-e_3}. \end{equation}
Likewise, we do not get anything new if we replace $e_1,e_2,e_3,e_4$ by $e_5,e_6,e_7,e_8$, since this exchange is equivalent
to replacing $s$ by $s''=s (x-e_5)(x-e_6)(x-e_7)(x-e_4)/y$.  So the choice of $\L$ depends only on the partition of the set $\{e_1,\dots,e_8\}$
as the union of two subsets $\{e_1,\dots,e_4\}$ and $\{e_5,\dots,e_8\}$, each with 4 elements.  There are 35 such partitions, and thus 35 even
spin structures with $h^0(\L)=0$.  They are obviously permuted transitively by permutations of the $e_a$.  So genus 3 surfaces with an
even spin structure of this type are parametrized by an irreducible divisor $\DD'\subset \M_{3,\spin+}$.

\subsection{Period Matrix Near $\DD$ and Near $\DD'$}\label{melfo}

 As described
in section \ref{twop}, mapping a super Riemann surface to the Riemann surface with the same period matrix
gives a meromorphic projection $\pi:\MM_{3,+}\to \M_{3,\spin+}$.  It is fairly obvious that $\pi$ has a pole along $\DD$, since the super period
matrix has a pole there.  It is less obvious that $\pi$ has a pole along $\DD'$; this was essentially explained to the author by P. Deligne.
In the remainder of this appendix, we have two goals. The first is to describe the origin of the pole of $\pi$ along $\DD'$.  The second
goal involves the following application to superstring perturbation theory.  Let $\Upsilon$ be the holomorphic measure on $\MM_{3,+}$ determined by superstring
theory on $\RR^{10}$.  
Integration over odd moduli of $\Sigma$ 
keeping fixing its  super period matrix $\h\Omega$ generates a natural measure $\pi_*(\Upsilon)$ on $\M_{3,\spin+}$.
Our second goal is to describe the behavior of $\pi_*(\Upsilon)$ along $\DD$ and along $\DD'$.  

We start by considering $\DD$.
We will need  to understand how the odd moduli of $\Sigma$ transform under the hyperelliptic ``involution'' $\tau$.
We have put the words ``involution'' in quotes because in acting on fermions, $\tau^2=-1$, not $+1$.  This is clear from the action of $\tau$ on $s=\sqrt{\d x/y}$.
Since $\tau$ acts  by $x,y\to x,-y$, it multiplies $\sqrt{\d x/y}$ by $\pm \sqrt{-1}$, where the choice of sign is arbitrary: it is up to us with which
sign we want to take $\tau$ to act on the spin bundle.  Let us make a choice of $\sqrt{-1}$ and declare that $\tau s=\sqrt{-1} s$.  Having made this
choice, it is now meaningful to ask how $\tau$ acts on the odd moduli of $\Sigma$.

It is convenient to write $K^{1/2}$ for what we have called $\L$ in Appendix \ref{spinstr} and $T^{1/2}$ for $\L^{-1}$.    We want to know how $\tau$
acts on $H^1(\Sigma_\red,T^{1/2})$, which parametrizes the odd moduli along the split locus.  It is slightly more convenient to determine
the action of $\tau$ on the dual space $H^0(\SIgma_\red,K^{3/2})$.  

Along $\DD$, the four sections of $K^{3/2}$ are $(\d x/y)^{3/2}x^t$, $t=0,\dots,3$.
Differently put, a general element of $H^0(\Sigma,K^{3/2}) $ is $s^3P_3(x)$ where $s=\sqrt{\d x/y}\in H^0(\Sigma_\red,K^{1/2})$ and $P_3(x)$ is a cubic polynomial.   $\tau$ acts as $-\sqrt{-1}$ on $s^3P_3(x)$, so dually it acts as $+\sqrt{-1}$ on $H^1(\Sigma_\red,T^{1/2})$.  In other words, $\Sigma$
has four odd moduli $\alpha_i$, $i=1,\dots,4$ all transforming as $\sqrt{-1}$ under $\tau$.

Along $\DD'$, the four sections of $K^{3/2}$ are 
\begin{equation}\label{firsttwo}\frac{\d x}{y} \sqrt{\frac{\d x}{y}(x-e_1)(x-e_2)(x-e_3)(x-e_4)}\,\cdot \,x^t,~~~ t=0,1,\end{equation}
transforming under $\tau$ as $-\sqrt{-1}$, 
and 
\begin{align}\label{secondtwo}&\frac{\d x}{y} \sqrt{\frac{\d x}{y}(x-e_1)(x-e_2)(x-e_3)(x-e_4)}\,\cdot \,\frac{y}{(x-e_1)(x-e_2)(x-e_3)(x-e_4)}\,\cdot\, x^t,~~~ t=0,1,\end{align}
transforming as $+\sqrt{-1}$.    Here $ \sqrt{\frac{\d x}{y}{(x-e_1)(x-e_2)(x-e_3)(x-e_4)}}$ is an abbreviation for $s(x-e_4)$, where $s$ was
characterized in 
eqns. (\ref{dundo}) and (\ref{Pundo}) and is assumed to transform as $\sqrt{-1}$.   (Note that $s(x-e_4)$ has simple zeroes at $e_1,\dots,e_4$, while $y/(x-e_1)(x-e_2)(x-e_3)(x-e_4)$ has
simple poles at those points, so that the two sections listed in eqn. (\ref{secondtwo}) are regular and nonzero at $e_1,\dots,e_4$, while the sections
in (\ref{firsttwo}) have simple zeroes there.  The behavior is reversed at $e_5,\dots,e_8$.  Also, $\d x/y$ is of order
$1/x^2$ at infinity, while $s(x-e_4)$ is of order $x$ at infinity and $y/(x-e_1)(x-e_2)(x-e_3)(x-e_4)$ is nonzero and bounded there; this is why
we take $t\leq 1$ in eqns. (\ref{firsttwo}) and (\ref{secondtwo}).)
Dual to this, along $\DD'$, $\Sigma$ has two odd moduli $\alpha_1,\alpha_2$ that transform as $\sqrt{-1}$ under $\tau$ and two odd moduli
$\beta_1,\beta_2$ that transform as $-\sqrt{-1}$.  

As we essentially learned in Appendix \ref{bosonic}, 
near either $\DD$ or $\DD'$, $\MM_{3,+}$ has   5 bosonic moduli
 $m_1,\dots,m_5$ that are even under $\tau$
and 1 bosonic modulus $\h\veps$ that is odd.  We define $\h\veps$ so that on the reduced space of $\MM_{3,+}$, it restricts to the parameter called $\veps$ in
Appendix \ref{bosonic}.  If we write $q$ for the same function of a $3\times 3$ period matrix that was introduced in section
\ref{bosonic}, then along the reduced space of $\MM_{3,+}$, we have $q=\h\veps^2$, precisely in parallel with the purely bosonic formula $q=\veps^2$.
However, the relation $q=\h\veps^2$ has corrections when the odd moduli are turned on.  Corrections to this relation that vanish at $\h\veps=0$
could be eliminated by redefining $\h\veps$, so we are only interested in corrections that are nonvanishing at $\h\veps=0$, or even have a pole
there.

First we consider the behavior along $\DD'$.  Since $q$ is a matrix element of the super period matrix, 
we can use the D'Hoker-Phong formula for the dependence of the super period matrix on odd moduli to compute its dependence on the
$\alpha_i $ and $\beta_j$.  
The leading correction to $q$ due to the odd moduli is a function bilinear in odd moduli. This function is constrained by
$\tau$-invariance, but is otherwise fairly generic, and in particular has no reason to vanish at $\h\veps=0$.  Thus, the bilinear correction to $q$
is $w_{ij}\alpha_i\beta_j$ for some $2\times 2$ matrix-valued function $w_{ij}$.
Generically along $\DD'$, this matrix is nondegenerate and we can pick the
odd moduli $\alpha_1,\alpha_2$ and $\beta_1,\beta_2$ so that 
\begin{equation}\label{trefo}q=\h\veps^2+\alpha_1\beta_1+\alpha_2\beta_2+\O(\alpha_1\alpha_2\beta_1\beta_2), \end{equation}
where the $\alpha_1\alpha_2\beta_1\beta_2$ term comes from the terms in the D'Hoker-Phong formula for the super period matrix that
are quartic in the odd variables.  
Since we have only explained what we mean by $\h\veps$ along the reduced space of $\MM_{3,+}$,
 we are free to redefine $\h\veps$ by adding holomorphic terms that are bilinear in the odd variables $\alpha_i, \,\beta_j$.
But it is not possible in this way to eliminate the nilpotent terms on the right hand side of eqn. (\ref{trefo}).

Along $\DD$, we have $H^0(\Sigma_\red,K^{1/2})\not=0$, and this means that the super period matrix has a pole at $\h\veps=0$;
the Dirac propagator, which enters the D'Hoker-Phong formula, is proportional to $\h\veps^{-1}$.  So the analog of eqn. (\ref{trefo})
is
\begin{equation}\label{prefo}q=\h\veps^2+\frac{\alpha_1\alpha_2+\alpha_3\alpha_4}{\h\veps}+\O\left(\frac{\alpha_1\alpha_2\alpha_3\alpha_4}{\h\veps^{2}}\right).\end{equation}
(In the D'Hoker-Phong expansion of the super period matrix, the term quartic in odd variables multiplies the product of two Dirac propagators and so is 
$\O(\h\veps^{-2})$.)

Eqn. (\ref{prefo}) makes clear that $\pi:\MM_{3,+}\to \M_{3,\spin+}$ has a pole along $\DD$.  Less obviously, however, this is
also true along $\DD'$.  This is because the natural variable that we use in expanding a string or superstring theory measure about the hyperelliptic
locus $\DD$ or $\DD'$ is not the matrix element $q$ of the period matrix, but its square root $\veps$, which is associated to a deformation of the
complex structure of $\Sigma_\red$ (or its metric); that is, it is associated to an element of $H^1(\Sigma_\red,T)$.   Rather than
give an abstract explanation of this statement, we refer the reader to Appendix \ref{benear}, where the point will hopefully become clear.
Regarding $\veps$ rather than $q=\veps^2$ as a local parameter along $\DD'\subset \M_{3,\spin+}$ amounts to treating
$\M_{3,\spin+}$ as a ``moduli stack'' rather than a ``moduli space.'' 

Let us see what happens if eqn. (\ref{trefo}) is written in terms of not $q$ but $\veps=q^{1/2}$.  We get
\begin{equation}\label{zrefo}\veps=\sqrt{\h\veps^2+\alpha_1\beta_1+\alpha_2\beta_2}+\dots =\h\veps+\frac{\alpha_1\beta_1+\alpha_2\beta_2}{2\h\veps}
+\dots.\end{equation}
The ellipses in eqn. (\ref{zrefo}) comes both from expanding the square root to higher orders and from the term quartic in odd variables
that was omitted in eqn. (\ref{trefo}).  
We should interpret this formula as giving the pullback of the function $\veps$ on $\M_{3,\spin+}$ to $\MM_{3,+}$:
\begin{equation}\label{erefo}\pi^*(\veps)= \h\veps+\frac{\alpha_1\beta_1+\alpha_2\beta_2}{2\h\veps}
+\dots.\end{equation}
Clearly, this pullback has a pole.  If we are supposed to take $\veps$ seriously as a function on $\M_{3,\spin+}$, then
the fact that it pulls back under $\pi$ to a function on $\MM_{3,+}$ with a pole along $\DD'$ shows that the projection
$\pi:\MM_{3,+}\to \M_{3,\spin+}$ has a pole along $\DD'$.

The analog of (\ref{zrefo}) along $\DD$ is
\begin{equation}\label{merox} \veps=\sqrt{\h\veps^2+\frac{\alpha_1\alpha_2+\alpha_3\alpha_4}{\h\veps}}=\h\veps+
\frac{\alpha_1\alpha_2+\alpha_3\alpha_4}{2\h\veps^2}+\dots. \end{equation}

\subsection{Behavior of $\pi_*(\Upsilon)$ Near $\DD$ and Near $\DD'$}\label{benear}

To describe the superstring measure $\Upsilon$ near $\DD'$, we use $\tau$-invariant local bosonic coordinates $\h m_i=\pi^*(m_i)$, along
with  bosonic and fermionic coordinates $\h\veps$ and $\alpha_1,\alpha_2,\beta_1,\beta_2$ that are not $\tau$-invariant.  The super Mumford
isomorphism (see \cite{YMa,BMFS, BS,RSVtwo} for original references and \cite{holomorphy} for an introduction) says that\footnote{The denominator
in this formula represents a section of
 the fifth power of the Berezinian of the cohomology of $\Sigma$.  To make eqn. (\ref{uerg}) simple and concrete, we have written in the
 denominator a section of this line
bundle over $\M_{3,\spin+}$.  To write an accurate formula, one should replace $\omega_t=x^t\,\d x /y$, $t=0,1,2$, with  corresponding
differentials on the super Riemann surface $\Sigma$.  The choice of these differentials will affect the function $F$, but has no essential
bearing on our discussion below.}
\begin{equation}\label{uerg} \Upsilon=\frac{F(\h\veps, \h m_1,\dots,\h m_5|\alpha_1,\alpha_2,\beta_1,\beta_2)[\d\h\veps \,\d \h m_1\cdots\d\h m_5|\d\alpha_1\d\alpha_2\d\beta_1\d\beta_2]}
{\left(\frac{\d x}{y}\wedge x\,\frac{\d x}{y}\wedge x^2\,\frac{\d x}{y}\right)^5}, \end{equation}
where the function $F$ is  holomorphic, nonzero, and $\tau$-invariant.  Notice that $\Upsilon$ is $\tau$-invariant: $\d\veps$ is odd under $\tau$, the denominator $\left(\frac{\d x}{y}\wedge x\,\frac{\d x}{y}\wedge x^2\,\frac{\d x}{y}\right)^5 $ is also odd, and the rest of the formula is $\tau$-invariant.    It is crucial that, despite being $\tau$-invariant, $\Upsilon$ cannot be written in terms of $\h\veps^2$ and $\d(\h\veps^2)$; as we will see momentarily, this fact
leads to a pole  in $\pi_*(\Upsilon)$ along $\DD'$.

To compute the measure $\pi_*(\Upsilon)$ on the reduced space $\M_{3,\spin+}$,
 the first step is to  re-express $\Upsilon$  in terms of $\veps$ rather than $\h\veps$.  We do this by solving eqn. (\ref{zrefo}) for $\h\veps$:
 \begin{equation}\label{zormo}\h\veps=\veps -\frac{\alpha_1\beta_1+\alpha_2\beta_2}{2\veps}+k(m_1,\dots,m_5)\frac{\alpha_1\alpha_2\beta_1\beta_2}{\veps^3}. \end{equation}
 The function $k(m_1,\dots,m_5)$ receives a contribution from the quartic terms that were omitted in eqn. (\ref{trefo}), and also from the expansion of the
 square root in eqn. (\ref{zrefo}).
 From (\ref{zormo}), we have
 \begin{equation}\label{telnox}\d\h\veps=-3k\frac{\d \veps}{\veps^4} \alpha_1\alpha_2\beta_1\beta_2+\dots,\end{equation}
 where we have written only the most singular term for $\veps\to 0$.  Making this substitution in (\ref{uerg}) and integrating over the odd variables,
 we find the singular behavior of $\pi_*(\Upsilon)$ along $\DD'$:
 \begin{equation}\label{really}\pi_*(\Upsilon)\sim \frac{-3k F}{\veps^4}\frac{ \d\veps \,\d m_1\dots\d m_5}{\left(\frac{\d x}{y}\wedge x\,\frac{\d x}{y}\wedge x^2\,\frac{\d x}{y}\right)^5}\end{equation}
 Thus $\pi_*(\Upsilon)$ is of order $1/\veps^4$ or $1/q^2$ along $\DD'$.
 
The literature actually contains a proposal \cite{CDvG} for a holomorphic measure on $\M_{3,\spin+}$ that is supposed to arise by integrating over the
odd variables in some fashion that has not been specified.  This formula is holomorphic along $\DD'$, so it does not coincide with $\pi_*(\Upsilon)$.
 
We can similarly determine the behavior of $\pi_*(\Upsilon)$ along $\DD$.   
The difference between $\DD$ and $\DD'$ is that, if $\Sigma$ is split, each of the 10 RNS fermions has a pair of zero-modes along $\DD$
and hence the path integral of the RNS fermions is proportional to $\veps^{10}$.  Along the split locus, therefore, one has near $\veps=0$
\begin{equation}\label{verg} \Upsilon=\veps^{10} \frac{G(\veps,m_1,\dots,m_5)
[\d\h\veps \,\d \h m_1\cdots\d\h m_5|\d\alpha_1\d\alpha_2\d\alpha_3\d\alpha_4]}
{\left(\frac{\d x}{y}\wedge x\,\frac{\d x}{y}\wedge x^2\,\frac{\d x}{y}\right)^5}, \end{equation} 
with $G$ non-zero.\footnote{At $\veps=0$, the cohomology $H^0(\Sigma,\mathit{Ber}(\Sigma))$ jumps, since $H^0(\SIgma_\red,K^{1/2})$
is nonzero along $\DD$.  The denominator in (\ref{verg}) therefore trivializes the appropriate line bundle only for $\veps\not=0$.  As a result
the formula (\ref{verg}), which vanishes at $\veps=0$, does not exhibit the super Mumford isomorphism at $\veps=0$.  To do so, one would have to write the formula in a more
sophisticated way, taking into account the jumping of the cohomology, but this is not necessary for our purposes, basically because there is no jumping
in the cohomology of $\Sigma_\red$.} To generalize this formula
away from the split locus, we have to take into account that, in conventional language, PCO insertions can absorb some zero-modes  of the matter
fermions, as discussed in section \ref{poleffect}.  However, the most singular behavior comes from expressing $\d\h\veps$ in terms of $\d\veps$.
The analog of eqn. (\ref{zormo}) is
\begin{equation}\label{wormo}\h\veps=\veps-\frac{\alpha_1\alpha_2+\alpha_3\alpha_4}{2\veps^2}+\frac{\alpha_1\alpha_2\alpha_3\alpha_4}{4\veps^5} .\end{equation}
(The $\alpha_1\alpha_2\alpha_3\alpha_4$ term comes from solving eqn. (\ref{merox}) for $\h\veps$ and does not depend on the 
$\O(\alpha_1\alpha_2\alpha_3\alpha_4/\h\veps^2)$ term in eqn. (\ref{trefo}).)
Hence
\begin{equation}\d\h\veps\sim-\frac{5}{4} \alpha_1\alpha_2\alpha_3\alpha_4 \frac{\d\veps}{\veps^6}+\dots, \end{equation}
where less singular terms are omitted.  Combining this factor of $\veps^{-6}$ with the $\veps^{10}$ in eqn. (\ref{verg}), and integrating over the odd
variables, we expect $\pi_*(\Upsilon)$
to vanish as $\veps^4=q^2$ along $\DD$.

\vskip 2cm\noindent
{\it Acknowledgments} Research was  partly supported by  NSF Grant PHY-1314311.  I would like to thank E. D'Hoker, R. Donagi, and D. Phong for discussions, and D'Hoker and Phong for help in reconciling some formulas
here with their results.  I also thank P. Deligne for detailed comments on
 an earlier version and for several helpful suggestions.

\bibliographystyle{unsrt}

\begin{thebibliography}{99}

\bibitem{RSV}
A. A. Rosly, A. S. Schwarz, and A. A. Voronov, ``Geometry Of Superconformal Manifolds,''
Comm. Math. Phys. {\bf 119} (1988) 129-152.



\bibitem{DPhagain}
E. D'Hoker and D. H. Phong, ``Conformal Scalar Fields And Chiral Splitting On Super Riemann Surfaces,'' Commun. Math. Phys. {\bf 125} (1989) 469-513.

\bibitem{DPhtwo}
E. D'Hoker and D. H. Phong, ``The Geometry Of String Perturbation Theory,'' Rev. Mod. Phys. {\bf 60} (1988) 917-1065.


\bibitem{Wittensurf}
E. Witten, ``Notes on Super Riemann Surfaces And Their Moduli,'' arXiv:1209.2459

\bibitem{DPh}
E. D'Hoker and D. H. Phong, ``Lectures On Two-Loop Superstrings,'' Adv. Lect. Math. {\bf 1} 85-123, hep-th/0211111.


\bibitem{DSW}
M. Dine, N. Seiberg, and E. Witten, ``Fayet-Iliopoulos Terms In String Theory,'' Nucl. Phys. {\bf B278} (1986) 769.

\bibitem{DIS}
M. Dine, I. Ichinose, and N. Seiberg, ``$F$ Terms And $D$ Terms In String Theory,''  Nucl. Phys. {\bf B293} (1987) 253.

\bibitem{ADS}
J. J. Atick, L. J. Dixon, and A. Sen, ``String Calculation Of 
Fayet-Iliopoulos $D$-Terms In Arbitrary Supersymmetric Compactifications,''
Nucl. Phys. {\bf B292} (1987) 109-149.

\bibitem{Wittenmore}
E. Witten, ``More On Superstring Perturbation Theory,''  arXiv:1304.2832.

\bibitem{DPhlatest}
E. D'Hoker and D. Phong,  ``Two-Loop Vacuum Energy For Calabi-Yau Orbifold Models,'' 
arXiv:1307.1749.

\bibitem{DPHcurrent}
E. D'Hoker and D. Phong, ``The Super Period Matrix With Ramond Punctures In The Supergravity Formulation,'' to appear. 


\bibitem{Martinec}
E. Martinec, ``Nonrenormalization Theorems And Fermionic String Finiteness,'' Phys. Lett. {\bf B171} (1986) 189-194.

\bibitem{Wittennotes}
E. Witten, ``Notes On Supermanifolds And Integration,'' arXiv:1209.2199.

\bibitem{WittenDonagi}
R. Donagi and E. Witten, ``Supermoduli Space is Not Projected,'' arXiv:1304.7798.

\bibitem{DM}
 P. Deligne and J. W. Morgan, ``Notes On Supersymmetry (following Joseph Bernstein),'' in P. Deligne et. al., eds., {\it Quantum Fields
And Strings: A Course For Mathematicians, Vol. 1} (American Mathematical Society, 1999).

\bibitem{Revisited}
E. Witten, ``Superstring Perturbation Theory Revisited,'' arXiv:1209.5461.

\bibitem{Verlindes}
E. Verlinde and H. Verlinde, ``Multi-Loop Calculations In Covariant Superstring Theory,''
Phys. Lett. {\bf B192} (1987) 95.

\bibitem{Sentwo}
J. J. Atick and A. Sen, ``Two-Loop Dilaton Tadpole Induced By Fayet-Iliopouplos D Terms
In Compactified Heterotic String Theory,'' Nucl. Phys. {\bf B296} (1988) 157-86.

\bibitem{YMa}
Yu. I. Manin, ``Critical Dimensions of  String Theories and the Dualizing Sheaf on the Moduli Space of (Super) Curves,'' Funct. Anal. Appl. {\bf 20} (1987) 244.

\bibitem{BMFS}
A. M. Baranov, Yu. I. Manin, I. V. Frolov, and A. S. Schwarz, ``A Superanalog Of The Selberg Trace Formula
And Multiloop Contributions For Fermionic Strings,'' Commun. Math. Phys. {\bf 111} (1987) 373-392.

\bibitem{BS}
M. A. Baranov and A. S. Schwarz, ``On The Multiloop Contribution To The String Theory,''
Int. J. Mod. Phys. {\bf A6} (1987) 1773-1796.

\bibitem{RSVtwo}
A. A. Rosly, A. S. Schwarz, and A. A. Voronov, ``Superconformal Geometry And String Theory,'' Commun. Math. Phys. {\bf 120} (1989) 
437-450.

\bibitem{holomorphy}
E. Witten, ``Notes On Holomorphic String And Superstring Theory Measures Of Low Genus,'' 
arXiv:1306.3621. 


\bibitem{CDvG}
S. L. Cacciatori, F. Della Piazza, and B. van Geemen, ``Modular Forms And Three Loop Superstring Amplitudes,'' arXiv:0801.2543,
 Nucl.Phys. B800 (2008) 565-590. 
 




\end{thebibliography}

\end{document}